\documentclass[aps,prd, reprint, superscriptaddress]{revtex4-2}

\usepackage{textcase}
\usepackage[T1]{fontenc}
\usepackage[utf8]{inputenc}
\usepackage{amsmath, amssymb}
\usepackage{graphicx}
\usepackage{hyperref}
\usepackage{geometry}
\usepackage{booktabs}
\usepackage{textcomp}
\usepackage{siunitx} % Optional: For better number/unit formatting if desired
\usepackage[table]{xcolor} % Required for \rowcolor (needs [table] option)

\begin{document}

\title{Thermal Noise Reduction in Ternary Optical Coatings: From Ti::GeO$_2$-Based Ternary Systems to High Index Materials.}
%: From Ti::GeO$_2$-Based Ternary Systems to High-Index Alternatives}
\author{Vincenzo Pierro}
\affiliation{Department of Engineering DING, University of Sannio, I-82100 Benevento, Italy}
\affiliation{INFN Sezione di Napoli, Gruppo Collegato di Salerno, I-84084 Fisciano (SA), Italy}

\author{Guerino Avallone}
%\email{guerino.avallone@sa.infn.it} % Esempio di email - SOSTITUISCI CON QUELLA REALE
% Oppure, se preferisci solo una nota senza email visibile (meno comune):
\thanks{corresponding author}
%\affiliation{INFN Sezione di Napoli, Gruppo Collegato di Salerno, I-84084 Fisciano (SA), Italy}
\affiliation{Dipartimento di Fisica "E. R. Caianello", Universit\`a di Salerno, I-84084 Fisciano (SA), Italy}

\author{Jessica Steinlechner}
\affiliation{Department of Gravitational Waves and Fundamental Physics, Maastricht University, P.O. Box 616, 6200 MD Maastricht, The Netherlands}

\author{Marco Bazzan}
\affiliation{Dipartimento di Fisica e Astronomia "Galileo Galilei", Universit\`a di Padova, I-35131 Padova, Italy}

\author{Francesco Chiadini}
\affiliation{Dipartimento di Ingegneria Industriale DIIN, Universit\`a di Salerno, I-84084 Fisciano (SA), Italy }
\affiliation{INFN Sezione di Napoli, Gruppo Collegato di Salerno, I-84084 Fisciano (SA), Italy}

\author{Roberta  De Simone}
\affiliation{Dipartimento di Ingegneria Industriale DIIN, Universit\`a di Salerno, I-84084 Fisciano (SA), Italy }
\affiliation{INFN Sezione di Napoli, Gruppo Collegato di Salerno, I-84084 Fisciano (SA), Italy}

\author{Marianna Fazio}
\affiliation{SUPA, School of Physics and Astronomy, University of Strathclyde, Glasgow G1 1XQ, United Kingdom}

\author{Massimo Granata}
\affiliation{Laboratoire des Matériaux Avancés (LMA), Univ Lyon, Universit\'{e} Claude Bernard Lyon 1, CNRS/IN2P3, F-69622 Villeurbanne, France}

\author{Veronica Granata}
\affiliation{Dipartimento di Ingegneria Industriale, Elettronica e Meccanica (DIIEM), Universit\`a degli Studi di Roma Tre, I-00146 Roma, Italy}
\affiliation{INFN Sezione di Napoli, Gruppo Collegato di Salerno, I-84084 Fisciano (SA), Italy}

\author{Gerardo Iannone}
\affiliation{INFN Sezione di Napoli, Gruppo Collegato di Salerno, I-84084 Fisciano (SA), Italy}
%\affiliation{Dipartimento di Fisica "E. R. Caianello", Universit\`a di Salerno, I-84084 Fisciano (SA), Italy}

\author{Graeme McGhee}
\affiliation{SUPA, School of Physics and Astronomy, University of Strathclyde, Glasgow G1 1XQ, United Kingdom}

\author{Carmen S Menoni}
\affiliation{Department of Electrical and Computer Engineering, Colorado State University, Fort Collins, Colorado 80523, USA}

\author{Christophe Michel}
\affiliation{Laboratoire des Matériaux Avancés (LMA), Univ Lyon, Universit\'{e} Claude Bernard Lyon 1, CNRS/IN2P3, F-69622 Villeurbanne, France}

\author{Vincenzo Fiumara}
\affiliation{Department of Engineering, University of Basilicata, I-85100 Potenza, Italy}
\affiliation{INFN Sezione di Napoli, Gruppo Collegato di Salerno, I-84084 Fisciano (SA), Italy}

% Ho espanso un po' l'affiliazione per chiarezza e standard.
%\author{Vincenzo Pierro$^{1,}$$^2$, Guerino Avallone$^{2,}$$^5$, Gerardo Iannone$^{2,}$$^5$, Veronica Granata$^{8,}$$^2$, \\ 
%Vincenzo Fiumara$^{7,2}$, Jessica Steinlechner$^3$, Marco Bazzan$^4$,  
%Christophe Michel$^6$\\, Massimo Granata$^6$, Carmen S Menoni$^9$, Marianna Fazio et al. \\
%{\small $^1$Department of Engineering DING at University of Sannio }\\
%{\small $^2$INFN di Napoli, Gruppo Collegato di Salerno}\\
%{\small $^3$Department of Gravitational Waves and Fundamental Physics at Maastricht University}\\
%{\small $^4$Dipartimento di Fisica e Astronomia "Galileo Galilei" at Universit\'{a} di Padova}\\
%{\small $^5$Dipartimento di Fisca "E. R. Caianello" at Universit\'{a} di Salerno}\\
%{\small $^6$Laboratoire des Matériaux Avancés L.M.A.- CNRS/IN2P3}\\
%{\small $^7$Department of Engineering of the University of Basilicata.}\\
%{\small $^8$Dipartimento di Ingegneria Industriale, Elettronica e Meccanica (DIIEM)}\\ {\small at Università degli Studi di Roma Tre}\\
%{\small $^9$Colorado State University, Fort Collins.}\\
%}

%\date{}

\begin{abstract}
Minimizing coating thermal noise is crucial for enhancing gravitational wave detector sensitivity, with a target Amplitude Spectral Density Reduction Factor (ASD RF) of $0.5$ relative to standard coatings. 
This study investigates the design of low-noise dielectric stacks using the 'Double Stack of Doublet' strategy, explored via ad-hoc optimization heuristics specifically developed for efficient parametric analysis
of coating performance.
We analyze the performance limits of ternary coatings based on SiO$_2$, Ti::SiO$_2$, and Ti::GeO$_2$, 
considering material property uncertainties and absorption constraints. Optimization results show that this system, even with relaxed absorbance constraint (1 ppm), falls short of the target, achieving a best ASD RF of $\sim 0.69$. 
Consequently, we explore alternative ternary 'Double Stack of Doublet' designs incorporating higher-refractive-index materials. 

Simulations demonstrate that incorporating alternative high-index materials offers a promising pathway, potentially enabling the achievement of the project target.
% Simulations demonstrate that utilizing dense materials as the third material offers a promising pathway, potentially achieving the project target. 
We discuss the optimization strategies, performance trade-offs, design robustness, and implications of using high-index, potentially higher-loss materials for next-generation 
optical coatings.
\end{abstract}

\maketitle

%%%%%%%%%%%%%%%%%%%%%%%%%%%%%%%%%%%%%%%%%%%%%%%%%%%%%%%%%%%%%
\section{Introduction}
%%%%%%%%%%%%%%%%%%%%%%%%%%%%%%%%%%%%%%%%%%%%%%%%%%%%%%%%%%%%%
The quest for increasing the sensitivity of gravitational wave detectors, such as Advanced  Virgo \cite{Virgosite}, LIGO \cite{LIGOsite}, and KAGRA \cite{KAGRA}, is fundamentally linked to the mitigation of limiting noise sources.
Among the most significant is coating thermal noise, particularly Brownian noise originating from the high-reflectivity multilayer dielectric coatings, deposited on the end test masses of the interferometer arms
\cite{HarryBook, Braginsky2002,Fejer2021LIGO,GV, Hong}.

Reducing this noise source is paramount for accessing weaker astrophysical signals and enabling new discoveries. In this frameowork, measuring it accurately is a crucial issue (see \cite{Pinto}).
The scientific community has set an ambitious target goal, often quantified by an Amplitude Spectral Density Reduction Factor (henceforth ASD RF) 
relative to reference coatings \cite{Granata2020cqg, testimone}, aiming for values as low as $0.5$ for next-generation detectors.
Achieving such low thermal noise reduction necessitates careful {\em Stack Engineering} – the meticulous design and optimization of the coating structure. This involves balancing intricate trade-offs between optical performance
(high reflectivity at the operational laser wavelength, typically $\lambda_0=1064$ nm), mechanical loss (quantified by the loss angle $\phi$ or the derived Braginsky coefficient $\eta$), 
and optical absorption ($\alpha_c$) related to the extinction coefficient $\kappa$.

This paper provides an update on our ongoing research line dedicated to designing these ultra-low noise coatings. We systematically analyze various coating configurations, moving from established binary systems to 
more complex ternary designs, with a particular focus on the  'Double Stack of Doublet' (henceforth DSD)  strategy, that is a particular form of multimaterial technolgy 
\cite{Multimat1,Multimat2,Multimat3,Multimat4}.
Our study evaluates the critical impact of material parameters, including their inherent uncertainties 
(as exemplified by the Braginsky coefficient  $\eta_{Ti::SiO_2}$ for titania-doped silica), on the achievable performance. We first investigate the performance limits of the  ternary system comprising Silica (SiO$_2$), 
Titania-doped Silica (Ti::SiO$_2$), and Titania-doped Germania (Ti::GeO$_2$), exploring different design constraints and optimization outcomes. Recognizing the limitations encountered with this 
system in reaching the ASD RF = 0.5 target, we then explore the potential of alternative strategies employing denser, higher-refractive-index materials, specifically Silicon Nitride (SiN$_x$) \cite{MGr}
and Amorphous Silicon (aSi) \cite{ASI}, as components in ternary DSD structures. Throughout this work, we utilize simulation, evolutionary optimization algorithms, and parametric analysis to map the design space, 
identify performance bottlenecks, and highlight promising pathways towards achieving the stringent requirements of future gravitational wave detectors.

This paper is structured as follows. We begin by outlining the noise models (Section II) and material parameter considerations, particularly for Ti::SiO$_2$
(Section III). Section IV evaluates the limits of binary coatings. We then investigate the ternary SiO$_2$/Ti::SiO$_2$/Ti::GeO$_2$
system, first by analyzing its fundamental absorption limits (Section V), and then by detailing the optimization and robustness of DSD designs under tight constraints (Section VI). 
An alternative strategy with relaxed absorption constraints for this system is explored in Section VII.
Subsequently, Section VIII assesses the potential of DSD designs incorporating denser, high-index materials (SiN$_x$, aSi).
Finally, Section IX presents our conclusions and discusses future directions. 
The Appendix provides a summary of Effective Medium Theory and a discussion of other relevant coating thermal noise models.

%%%%%%%%%%%%%%%%%%%%%%%%%%%%%%%%%%%%%%%%%%%%%%%%%%%%%%%%%%%%%
\section{The noise model}
%%%%%%%%%%%%%%%%%%%%%%%%%%%%%%%%%%%%%%%%%%%%%%%%%%%%%%%%%%%%%

\begin{figure}[t] % Opzioni di posizionamento: qui, top, bottom, pagina di floats
  \centering % Centra la figura orizzontalmente
  \includegraphics[width=0.4\textwidth]{"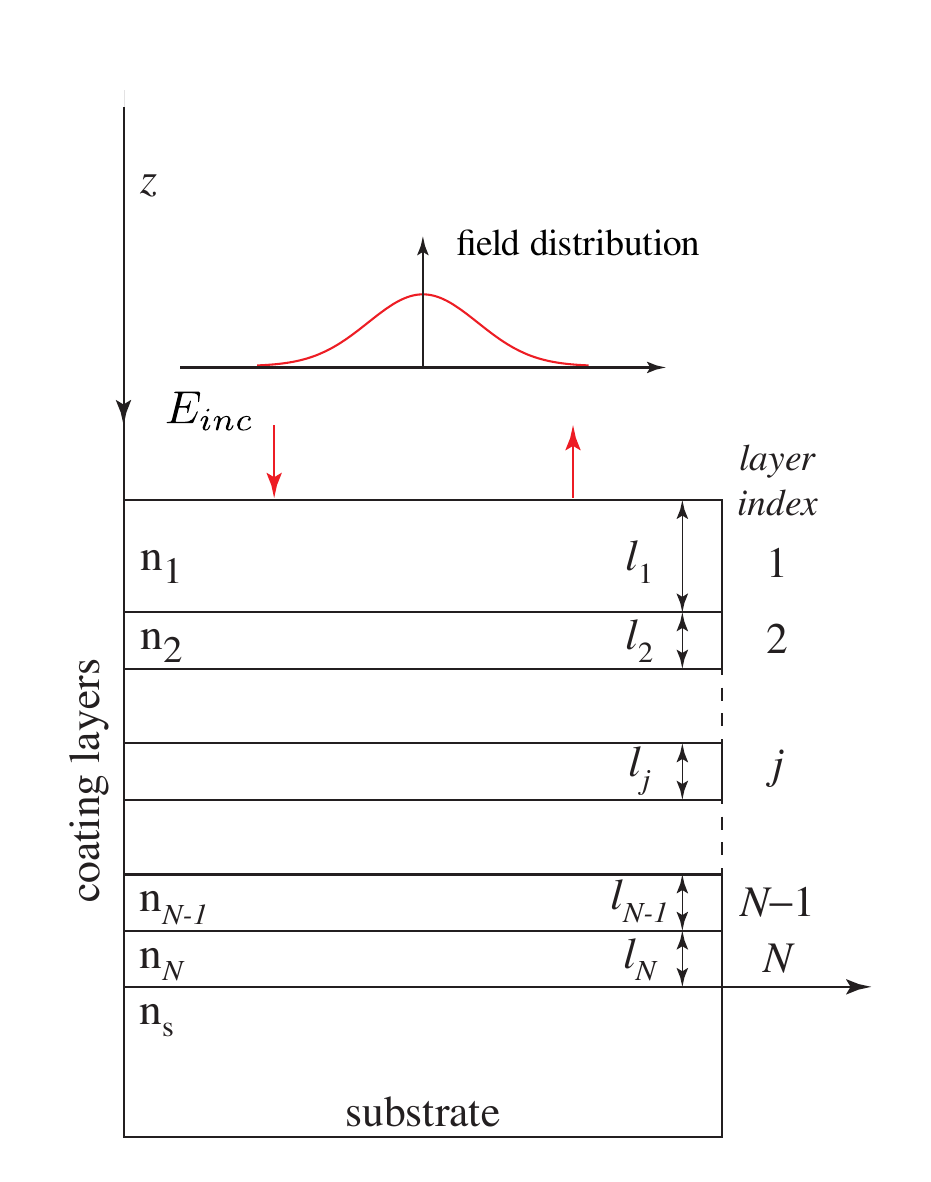"} % Includi l'immagine (sostituisci con il tuo file)
  \caption{A typical coating made of $N$ layers deposited on a substrate. 
  The Electromagnetic field (Gaussian beam shaped) is incident form above (vacuum half space). In the plane wave approximation the field phasor is $E_{inc} \exp{i \omega t}$ where the angular frequency
  corresponds to the free space wavelength $\lambda_0=1064$ nm.}
  \label{fig:model}
\end{figure}

The coating is a layered structure made of planar layers in principle of different materials and thicknesses, as depicted in Fig. \ref{fig:model}. The electromagnetic field (with a gaussian shape, and wavelength $\lambda_0=1064$ nm) 
is incident from the above. In the following we make the plane wave approximation to develop optical simulations (by using the classical matrix formalism \cite{Abele}), 
in this context the harmonic field is described by the  $E_{inc} \exp(i \omega t)$ where $\omega$ is the angular frequency corresponding to the laser operating free space wavelength $\lambda_0= 1064$ nm. 

To understand and mitigate thermal noise in optical coatings, which is critically important in precision measurements and applications like gravitational wave interferometers, several models have been developed. 
The \textit{Braginsky model} \cite{HarryBook} provides a fundamental framework for describing the coating Brownian noise power spectral density ($S_{CB,j}$) of a single layer $j$ as:
\begin{widetext}
\begin{equation}
    S_{CB,j} = \frac{2k_B T \phi'_j l_j}{\pi^2 w_m^2 f} \left[ \frac{(1 + \sigma'_j)(1 - 2\sigma'_j)}{Y'_j(1 - \sigma'_j)} + \frac{Y'_j(1 + \sigma)^2(1 - 2\sigma)^2}{Y^2(1 - \sigma_j^2)} \right],
    \label{eq:BGV}
\end{equation}
\end{widetext}
where $k_B$ is Boltzmann's constant, $T$ is the absolute temperature,  $w_m$ relates to the laser beam size, and $f$ is the frequency.
Furthermore in eq. (\ref{eq:BGV}) we have the layer's specific quantities:  the mechanical loss coefficient $\phi'_j$,  the Young modulus $Y'_j$, the Poisson ratio $\sigma'_j$ ,  and  the thickness $l_j$ .
 The term $\eta_j$ 
 \begin{widetext}
\begin{equation}
    \eta_j = \phi'_j Y \left[ \frac{(1 + \sigma'_j)(1 - 2\sigma'_j)}{Y'_j(1 - \sigma'_j)} + \frac{Y'_j(1 + \sigma)^2(1 - 2\sigma)^2}{Y^2(1 - \sigma_j^2)} \right].
    \label{eq:BGVcoef}
\end{equation}
\end{widetext}
is often defined to simplify the expression (\ref{eq:BGV}), representing the material-dependent part of the thermal noise, 
this coefficient (\ref{eq:BGVcoef}) depend also on Young’s modulus $Y$ and the Poisson’s ratio $\sigma$ of the substrate. 

For multilayer coatings, a linear model approximates the total noise ($S_{CB}$) by summing the contributions from each layer ($N_L$ layers):
\begin{equation}
    S_{CB} = \sum_{j=1}^{N_L} S_{CB,j} = \frac{2k_B T}{\pi^2 w_m^2 f Y} \sum_{j=1}^{N_L} \eta_j l_j.
\end{equation}
It is well known that the {\em Fejer Effective Medium Theory} offers a more generalized approach \cite{Fejer2021LIGO}, (see also the supplementary material of \cite{Vajente}), considering both bulk and shear loss components to provide a generalized version of formula (\ref{eq:BGVcoef}) 
and a more accurate description of thermal noise, especially in complex multilayer structures (see Appendix).  
In spite of the greater accuracy of the physical model, Fejer's formula reduces to Braginsky's in the approximation that bulk and shear mechanical losses are equal (a condition that applies to almost all materials considered,
 see Appendix).  
These models highlight the importance of material properties like mechanical loss ($\phi$), Young's modulus ($Y$), Poisson ratio ($\sigma$), and refractive index in minimizing thermal noise in advanced optical systems.

%%%%%%%%%%%%%%%%%%%%%%%%%%%%%%%%%%%%%%%%%%%%%%%%%%%%%%%%%%%%%
%\section{Mechanical Parameters and Braginsky Coefficient}
%%%%%%%%%%%%%%%%%%%%%%%%%%%%%%%%%%%%%%%%%%%%%%%%%%%%%%%%%%%%%

Here, we consider the doped Silica material in a possible ternary configuration Ti::SiO$_2$/ SiO$_2$/ Ti::GeO$_2$. 

Ternary coating designs, such as those employing Silica, doped-Silica, and Ti::GeO$_2$, are emerging as promising alternatives to binary coatings  (e.g. SiO$_2$/Ti::Ta$_2$O$_5$ \cite{testimone} and SiO$_2$/Ti::GeO$_2$ \cite{Vajente} ) 
for advanced optical applications requiring very low thermal noise. 
 
These ternary systems offer a potentially finer degree of control over the refractive index profile and material properties within the coating stack. 
This approach could lead to improved impedance matching between layers, potentially reducing interface losses and scattering. Furthermore, in comparison to other ternary designs like
 Silica/Ti::Ta$_2$O$_5$/Ti::GeO$_2$ coatings, the Silica/Ti::Silica/Ti::GeO$_2$ combination might offer advantages in terms of material compatibility and stress management, given the 
 compositional similarity within the silica-based components.  
While both ternary strategies aim to optimize coating performance beyond binary systems, the specific merits of Silica/Ti::Silica/doped-GeO$_2$ coatings warrant further investigation as a potentially advantageous pathway for minimizing thermal noise in high-precision optical systems.

%%%%%%%%%%%%%%%%%%%%%%%%%%%%%%%%%%%%%%%%%%%%%%%%%%%%%%%%%%%%%%\section{Braginsky Coefficient for Ti::SiO$_2$}

The best-known composition range for Ti::SiO$_2$ varies between $50$\% and $70$\% \cite{LIGOdoc}. 
Let us consider the normalized Braginsky coefficient $\bar{\eta}_j$, useful in the optimization algorithm, defined by:
\begin{equation}
    \bar{\eta}_j = \frac{\eta_j}{\eta_{SiO_2}}.
\label{eq:quattro}
\end{equation}
where index $j$ specifies the coating material.
We define a normalization with respect to the Braginsky coefficient of silica ($\eta_{SiO_2}$),  because silica is the material (at room temperature) with the lowest known thermal noise and can be used as a benchmark.
The value of Eq. (\ref{eq:quattro}) can be computed by using the data for mechanical loss coefficient $\phi$ at $100$ Hz, Young’s modulus $Y$, 
and Poisson’s ratio $\sigma$ given in \cite{LIGOdoc, TiSilica, Netterfield2007}.

The mechanical loss in thin-film coatings is a critical parameter for the design of high-precision optical systems. 
The data in  \cite{TiSilica} proposed a measured value for the mechanical loss angle of Ti::SiO$_2$, yielding $\phi_{Ti:SiO_2} = 1.44 \times 10^{-4}$.

Table \ref{tab:braginsky_coeff} presents the relevant mechanical properties for different coating materials to compute formula (\ref{eq:quattro}), 
 used in the following of the paper for all simulations.
\begin{table*}[t]
    \centering
    \caption{Mechanical and optical properties  for different materials and substrate \cite{Granata2020cqg, LMA}. 
    The available values for doped silica are affected by uncertainty because this material has only recently been investigated}
    \vspace{0.3cm}
    \begin{tabular}{lccccc}
        \hline
        Material & $n_r$ @ 1064 nm \,\,& $\kappa$ @ 1064 nm & $Y$ (GPa) & $\sigma$ & $\phi$ [rad] @ 100 Hz \\
        \hline
        SiO$_2$ (Suprasil\texttrademark) &1.4499 & $\le 10^{-8}$ & 73  & 0.17 & $<1 \times 10^{-8}$ \\
        SiO$_2$ & 1.45 & $3\times 10^{-8}$ & 70  & 0.19 & $2.3 \times 10^{-5}$ \\
        Ti::SiO$_2$ & [1.92,1.97] & $\le 2\times 10^{-7}$ & [85.6, 110]  & [0.3,0.4] & $1.44 \times 10^{-4}$ \\
        Ti::GeO$_2$  & $1.89$ & $1.9 \times 10^{-7}$ & 92  & 0.25 & $1 \times 10^{-4}$ \\
        \hline
    \end{tabular}
    \label{tab:braginsky_coeff}
\end{table*}

%%%%%%%%%%%%%%%%%%%%%%%%%%%%%%%%%%%%%%%%%%%%%%%%%%%%%%%%%%%%%%%%%%%%%%%%%%%%%%%%%%%%%%%%%%%%%%%%%%%%%%%%%%%%%%%%%%%%%%%%%%%%%%%%%%%
\section{Distribution of the $\bar{\eta}_{Ti::SiO_2}$ Coefficient}
%%%%%%%%%%%%%%%%%%%%%%%%%%%%%%%%%%%%%%%%%%%%%%%%%%%%%%%%%%%%%%%%%%%%%%%%%%%%%%%%%%%%%%%%%%%%%%%%%%%%%%%%%%%%%%%%%%%%%%%%%%%%%%%%%%%

The mechanical properties of silica and doped tantala, traditional materials for gravitational interferometer coatings, are well-established through extensive experimental research \cite{{Granata2020cqg}}. 
Conversely, doped silica has only recently been explored for this role. Therefore, its mechanical parameters are characterized by greater variability.

The distribution of measured Poisson’s ratio versus measured Young’s modulus for a fixed $\phi_{Ti::SiO_2}$ value shows significant data dispersion, introducing uncertainty in determining $\eta_{Ti::SiO_2}$. 
The value distribution suggests a need for further refinement in material characterization.

\begin{figure}[h] % Opzioni di posizionamento: qui, top, bottom, pagina di floats
%  \centering % Centra la figura orizzontalmente
 \hspace*{-0.5cm} \includegraphics[width=0.45\textwidth]{"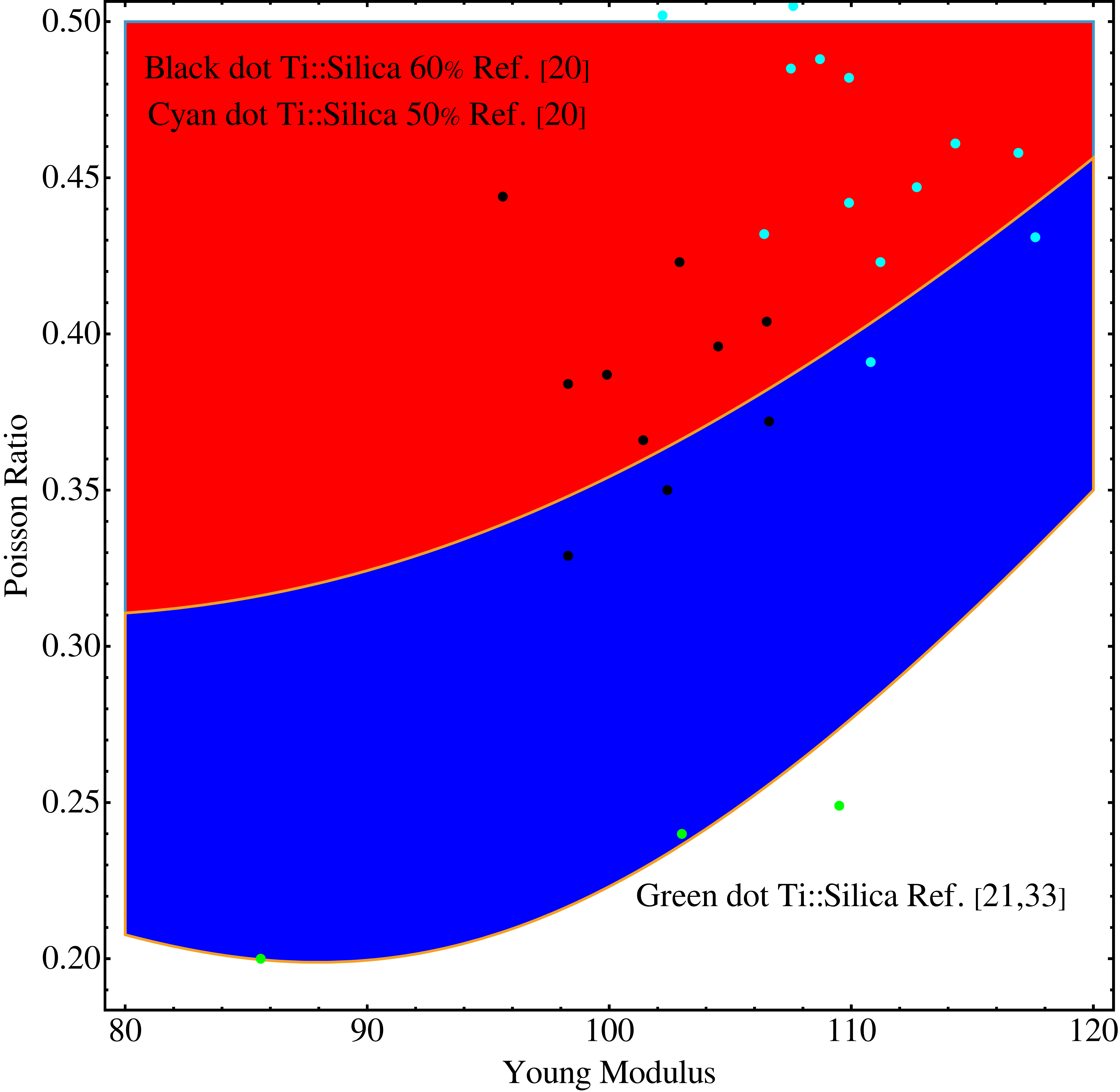"} % Includi l'immagine (sostituisci con il tuo file)
  \caption{Contour plot of the normalized Braginsky coefficient as a function of Young's Modulus and Poisson's Ratio, for a fixed mechanical loss factor $\phi_{Ti:SiO_2}=1.44 \times 10^{-4}$.
   Black and cyan dots are data points from LIGO document G2001684 \cite{LIGOdoc}. The green dots specifically represent measurements from the Glasgow group \cite{TiSilica,GraemePri}. 
   The red region corresponds to values of $\bar{\eta}_{Ti::SiO_2} < 5.6$, while the blue region corresponds to values of $5.6 < \bar{\eta}_{Ti::SiO_2} < 6.7$.}
  \label{fig:fig1}
\end{figure}

\begin{figure}[h] % Opzioni di posizionamento: qui, top, bottom, pagina di floats
%  \centering % Centra la figura orizzontalmente
 \hspace*{-0.5cm} \includegraphics[width=0.5\textwidth]{"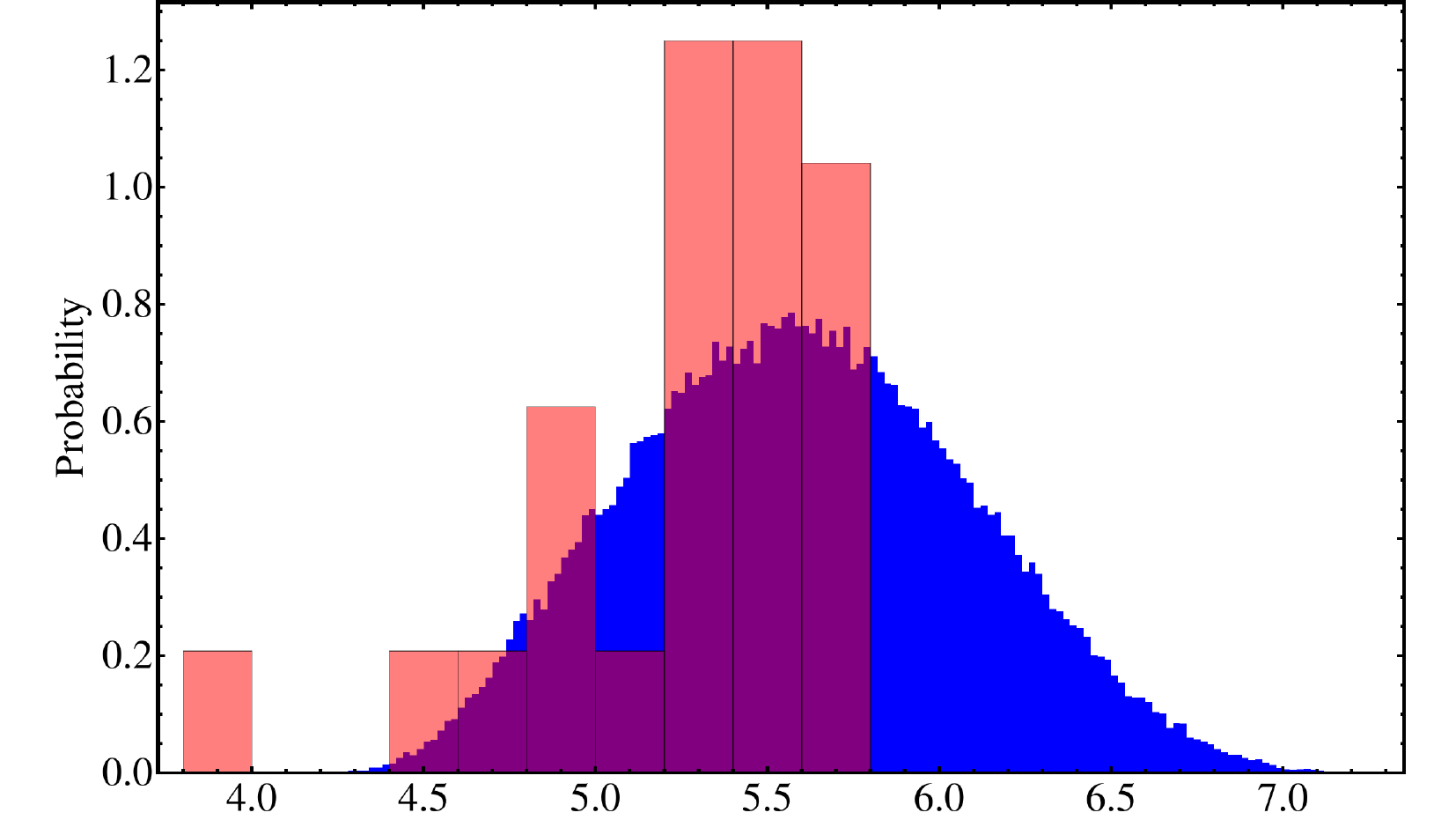"} % Includi l'immagine (sostituisci con il tuo file)
  \caption{This figure presents an uncertainty analysis of the normalized Braginsky coefficient ($\eta_{Ti::SiO_2}$) for Ti::Silica coatings, assuming a mechanical loss value of $\phi_{Ti:SiO_2} = 1.44 \times 10^{-4}$. 
  The pink histogram illustrates the distribution of $\bar{\eta}_{Ti::SiO_2}$ derived from experimental data.
   The blue distribution is obtained by considering the uncertainties in Young’s Modulus ($Y_{Ti::SiO_2}$) and Poisson’s ratio ($\sigma_{Ti::SiO_2}$),
    assuming uniform distributions within the ranges of $\sigma_{Ti::SiO_2} = [0.3, 0.4]$ and $Y_{Ti::SiO_2} = [85.6, 110]$ GPa.}
  \label{fig:fig2}
\end{figure}

The two graphs in Fig.s \ref{fig:fig1} and  \ref{fig:fig2} present an analysis of the elastic parameters and thermal noise properties of Ti::SiO$_2$. 
The graph in Fig. \ref{fig:fig1} shows the distribution of Young’s modulus and Poisson’s ratio, 
with colored regions distinguishing specific ranges of the parameter \(\bar{\eta}_{Ti::SiO_2}\). 
Black and cyan dots correspond to different experimental compositions, while the red and blue regions indicate \(\bar{\eta}_{Ti::SiO_2} \leq 5.6\) and \(5.6 \leq \bar{\eta}_{Ti::SiO_2} \leq 6.07\), respectively. 

The graph in Fig. \ref{fig:fig2} displays the probability distribution of \(\bar{\eta}_{Ti::SiO_2}\), obtained using experimental values of \(\phi_{Ti::SiO_2}\), \(\sigma_{Ti::SiO_2}\), and \(Y_{Ti::SiO_2}\) based on availabe measurement. 
The estimated mean value of \(\bar{\eta}_{Ti::SiO_2}\) is $5.6$, with a standard deviation of approximately $0.49$, while the three-standard-deviation confidence interval ranges from $4.13$ to $7.07$. 
% These results suggest good consistency with experimental data and help identify suitable parameter ranges for specific applications.

The uncertainty analysis illustrated in  Fig. \ref{fig:fig2} was conducted considering the mechanical loss the value $\phi_{Ti:SiO_2} = 1.44 \times 10^{-4}$
In more details the pink histogram represents the distribution of $\bar{\eta}_{Ti::SiO_2}$ derived from experimental values reported in \cite{LIGOdoc, TiSilica, Netterfield2007} .
The distribution of $\eta_{Ti::SiO_2}$ is derived by considering the uncertainties in 
Young's Modulus ($Y_{Ti::SiO_2}$) and Poisson's ratio ($\sigma_{Ti::SiO_2}$), assuming uniform distributions within the ranges shown: $\sigma_{Ti::SiO_2} = [0.3, 0.4]$ and $Y_{Ti::SiO_2} = [85.6, 110]$ GPa.  
The resulting distribution of $\bar{\eta}_{Ti::SiO_2}$ is approximately Gaussian, with a mean value of $5.6$ and a standard deviation of $0.491$, as indicated in the figure.

Based on both Fig. \ref{fig:fig1} and Fig. \ref{fig:fig2}, it is evident that current measurements lead to a significant dispersion in the values of $\bar{\eta}_{Ti::SiO_2}$.  
While our estimation provides a reasonable assessment, even more pessimistic estimations of $\phi_{Ti::SiO_2}$ exist, which would result in $\bar{\eta}_{Ti::SiO_2}$ values too high to be practically 
useful in a ternary system.
In order to reduce uncertainty, it would be more advantageous to directly measure $\bar{\eta}_{Ti::SiO_2}$ (e.g. using Coating Thermal Noise interferometer \cite{CTN}) 
rather than relying on separate measurements of $\phi_{Ti::SiO_2}$, $\sigma_{Ti::SiO_2}$, and $Y_{Ti::SiO_2}$.

%%%%%%%%%%%%%%%%%%%%%%%%%%%%%%%%%%%%%%%%%%%%%%%%%%%%%%%%%%%%%%%%%%%%%%%%%%%%%%%%%%%%%%%%%%%%%%%%%%%%%%%%%%%%%%%%%%%%%%%%%%%%%%%%%%%
\section{Binary Coating Silica/doped-Silica}
%%%%%%%%%%%%%%%%%%%%%%%%%%%%%%%%%%%%%%%%%%%%%%%%%%%%%%%%%%%%%%%%%%%%%%%%%%%%%%%%%%%%%%%%%%%%%%%%%%%%%%%%%%%%%%%%%%%%%%%%%%%%%%%%%%%

The coating design aims to minimize ASD RF while satisfying the constraints on transmittance at the coating-cavity interface ($\tau_c \le 6$ ppm) and coating absorbance ($\alpha_c \le 0.5$ ppm) \cite{Nota}.
Different stack configurations are evaluated in terms of noise minimization. 
At first, binary  coatings made of Silica and doped-Silica are studied to identify the limit in absorption and transmittance.

In the following paragraphs, we will utilize refractive index values measured in \cite{TiSilica} $n_r=1.92$ for the doped Silica. 
We will make realistic assumptions about the extinction coefficient (imaginary part of the refractive index) and establish usability limits for the doped silica based on these considerations.

\begin{figure}[h] % Opzioni di posizionamento: qui, top, bottom, pagina di floats
\centering % Centra la figura orizzontalmente
\hspace*{-0.5cm}\includegraphics[width=0.5\textwidth]{"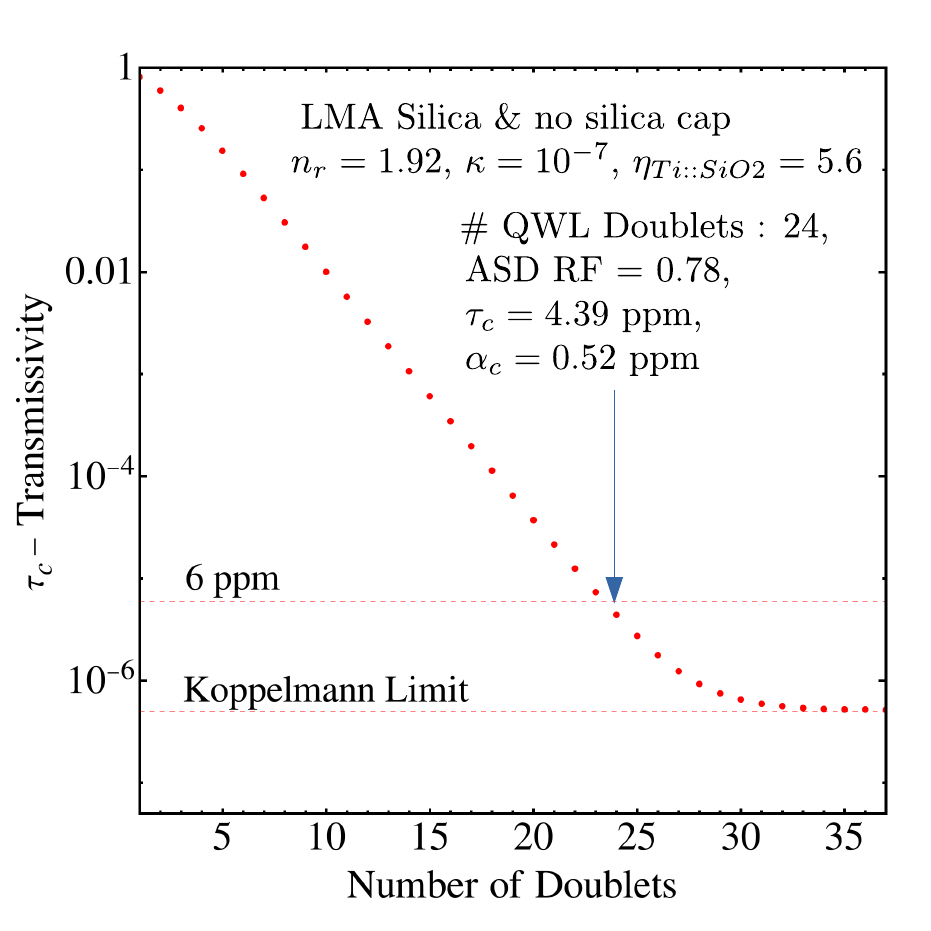"} % Includi l'immagine (sostituisci con il tuo file)
\caption{The transmissivity $\tau_c$ (or transmittance $\tau_c=1-|\Gamma|^2$) as a function of the number of doublets for QWL design. Here $\Gamma$
is the reflection coefficient at vacuum interface. 
Simulation parameters are relative to LMA Silica coating; the configuration does not include a half-wave cap.
The labels placed above the horizontal red dashed grid lines explain what each line represents.}
\label{fig:fig3}
\end{figure}

The Fig. \ref{fig:fig3} illustrates the transmissivity ($\tau_c$) of a binary SiO$_2$/Ti::SiO$_2$ Quarter-Wave Layer (QWL) coating as a function of the number of doublets.  
An increase in the number of doublets leads to both a reduction in coating transmissivity and an increase in thermal noise.

The figure highlights the {\em Koppelmann Limit} (henceforth KL)  at $0.51$ ppm. This value, theoretically derived by Koppelmann (see \cite{Koppelmann}), 
represents a fundamental lower bound for the transmissivity (and consequently for the absorption) achievable with this type of binary coating design. 
Our numerical simulations, conducted with custom-developed codes, confirm this theoretical prediction.

A QWL design with $24$ doublets meets the required transmittance, while its absorption reaches the KL.
The calculation is based on parameters masured by LMA team \cite{LMA}, with a refractive index $n_r = 1.92$, an extinction coefficient 
$\kappa = 10^{-7}$, and a Braginsky coefficient $\eta_{Ti::SiO_2} = 5.6$. The resulting transmissivity $\tau_c$ is $4.39$ ppm, and the absorption coefficient $\alpha_c$ is $0.52$ ppm for a design with 
$24$ doublets, achieving an ASD RF of $0.78$.

\begin{table*}[t]
\centering
\caption{Simulation Results: {\em Koppelmann limit} (KL) for QWL designs}
\begin{tabular}{@{}llllll@{}}
\toprule
Index (Re + Im) & BGV coef & KL & $\tau_c$ [ppm] & \# QWL Doublets & ASD RF \\
\midrule
$1.92 - 1.0 \times 10^{-7}i$ & 5.6 & 0.5157 & 4.39 & 24 & 0.7842 \\
$1.92 - 1.0 \times 10^{-7}i$ & 6.0 & 0.5157 & 4.39 & 24 & 0.8067 \\
$1.92 - 1.5 \times 10^{-7}i$ & 5.6 & 0.7140 & 4.58 & 24 & 0.7842 \\
$1.92 - 1.5 \times 10^{-7}i$ & 6.0 & 0.7140 & 4.58 & 24 & 0.8067 \\
$1.92 - 2.0 \times 10^{-7}i$ & 5.6 & 0.9124 & 4.78 & 24 & 0.7842 \\
$1.92 - 2.0 \times 10^{-7}i$ & 6.0 & 0.9124 & 4.78 & 24 & 0.8067 \\
\bottomrule
\end{tabular}
\label{tab:tab2}
\end{table*}

In order to perform a parametric analysis of the dependence of the KL on the extinction coefficient of doped silica, we report the results in Table \ref{tab:tab2},
which presents simulation results for QWL designs, varying the complex refractive index and the Braginsky coefficient of doped silica (assuming both realistic and pessimistic case for $\eta_{Ti::SiO_2}$).
 
For each parameter set, the table shows the KL and the achievable reduction factor of ASD (Amplitude Spectral Density), indicated by the ASD RF column.
We deduce that an extinction value (imaginary refractive index) of  $10^{-7}$ or less is crucial to achieve the prescribed $0.5$ ppm absorption limit for next-generation GW interferometers.
In any case, results in Table \ref{tab:tab2} show that even assuming an optimistic value for extinction, the achievable reduction factor is $0.78$, which is far above the design limit.

\begin{figure}[h] % Opzioni di posizionamento: qui, top, bottom, pagina di floats
\centering % Centra la figura orizzontalmente
\hspace*{-0.7cm}\includegraphics[width=0.55\textwidth]{"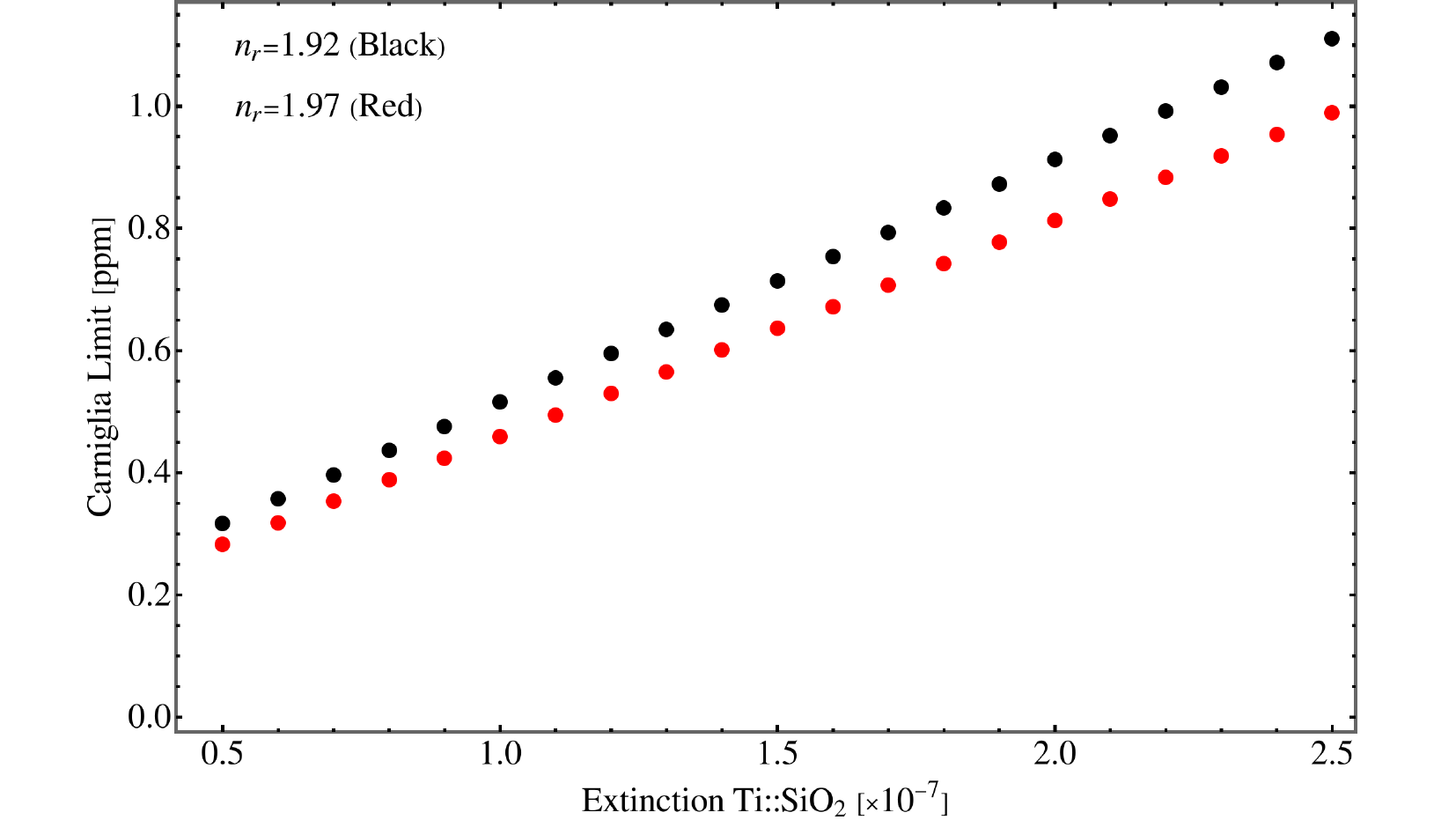"} % Includi l'immagine (sostituisci con il tuo file)
\caption{The {\em Carniglia's Limit} [ppm] for a binary coating of silica and doped silica, plotted as a function of the extinction coefficient of the doped silica (Ti:SiO$_2$).
 The two datasets correspond to different assumed real refractive indices for the doped silica layer: $n_r = 1.92$ (black dots) and $n_r = 1.97$ (red dots).}
\label{fig:fig4}
\end{figure}

The ultimate limit in absorbance for a binary coating made by Silica and doped-Silica is further elucidated in Fig. \ref{fig:fig4}, where is shown the {\em Carniglia's Limit} \cite{carniglia} (henceforth CL) for this two materials.

It is worth noting that the KL generally applies specifically to quarter-wave layer (QWL) designs. 
The CL, conversely, offers a more general bound on absorbance, valid regardless of the individual layer thicknesses. 
For this reason, the CL is typically lower than or equal to the KL. 
However, in the specific case examined here, utilizing materials (silica and doped silica) with extremely low extinction coefficients, the numerical values of the two limits are very close. 
Their difference typically becomes apparent only at the third decimal place or beyond.

% In Appendix B the question of the ultimate limit in absorbance of a binary coating is studied in detail for two index values of doped silica, $n_r=1.92$ (realistic) and $n_r=1.97$ (optimistic).

%%%%%%%%%%%%%%%%%%%%%%%%%%%%%%%%%%%%%%%%%%%%%%%%%%%%%%%%%%%%%%%%%%%%%%%%%%%%%%%%%%%%%%%%%%%%%%%%%%%%%%%%%%%%%%%%%%%%%%%%%%%%%%%%%%%
\section{Ternary Coating: {{ Silica/Ti::SiO$_2$/Ti::GeO$_2$}} Carniglia's limit.}
%\section{Ternary Coating: Silica/T\MakeLowercase{i}::S\MakeLowercase{i}O$_2$/T\MakeLowercase{i}::G\MakeLowercase{e}O$_2$ {\em Carniglia's limit}.}
%%%%%%%%%%%%%%%%%%%%%%%%%%%%%%%%%%%%%%%%%%%%%%%%%%%%%%%%%%%%%%%%%%%%%%%%%%%%%%%%%%%%%%%%%%%%%%%%%%%%%%%%%%%%%%%%%%%%%%%%%%%%%%%%%%%

In this section, we extend the concept of the CL, originally formulated for binary coating systems, to a more complex ternary structure composed of two overlapping stack made of doublets. 
The first stack is formed by alternating layers of silica and doped silica, while the second consists of silica and doped GeO$_2$ material. 
This ternary design framework allows for a richer parameter space, enabling us to explore how the minimal achievable absorbance is influenced by the extinction coefficients and refractive index of doped-Silica. 
The primary goal of this analysis is to identify the threshold value for the extinction coefficient of doped silica that still allows the system to reach a total absorbance as low as $0.5$ ppm. 

To efficiently explore the design space of the ternary optical coatings, we developed a 
fast heuristic computational method based on the Borg-MOEA evolutionary optimization framework \cite{BORG} (see also \cite{BORGVP} for specific application to gravitational wave interferometer cavity mirrors).

\begin{figure}[h] % Opzioni di posizionamento: qui, top, bottom, pagina di floats
\centering % Centra la figura orizzontalmente
\hspace*{-0.3cm}\includegraphics[width=0.47\textwidth]{"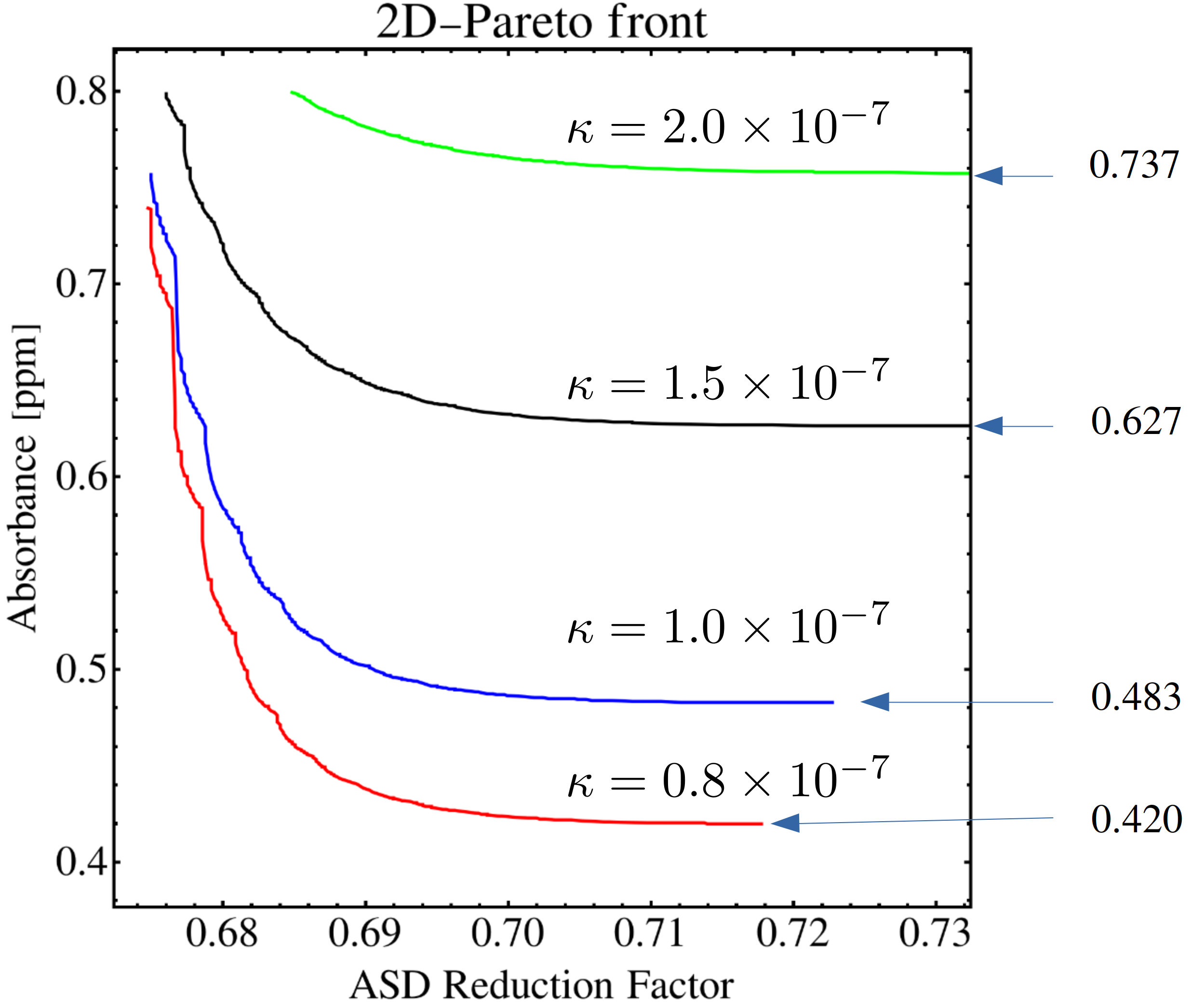"} % Includi l'immagine (sostituisci con il tuo file)
\caption{Pareto front Absorption versus ASD RF for different values of doped-Silica extinction $\kappa$. 
The refractive index is $n_r=1.92$ and the maximum number of doublets is equal to $40$. The numerical value of the stack limit absorbance is
displayed on the right near the arrows (for each curve).}
\label{fig:fig5}
\end{figure}

The algorithm employs the thicknesses of the individual layers and the number of doublets in the first and second stacks as parameters on which to perform the optimization.
This multi-objective evolutionary algorithm is well-suited for identifying trade-offs between conflicting design goals, in this case absorption and noise reduction factor, and was employed to compute the Pareto front of the system. 
 The results of these simulations, showcasing the optimal configurations and their corresponding performance metrics, are presented in Fig. \ref{fig:fig5} for different values of Ti::SiO$_2$ extinction coefficient.

\begin{figure}[h] % Opzioni di posizionamento: qui, top, bottom, pagina di floats
\centering % Centra la figura orizzontalmente
\hspace*{-0.7cm}\includegraphics[width=0.4\textwidth]{"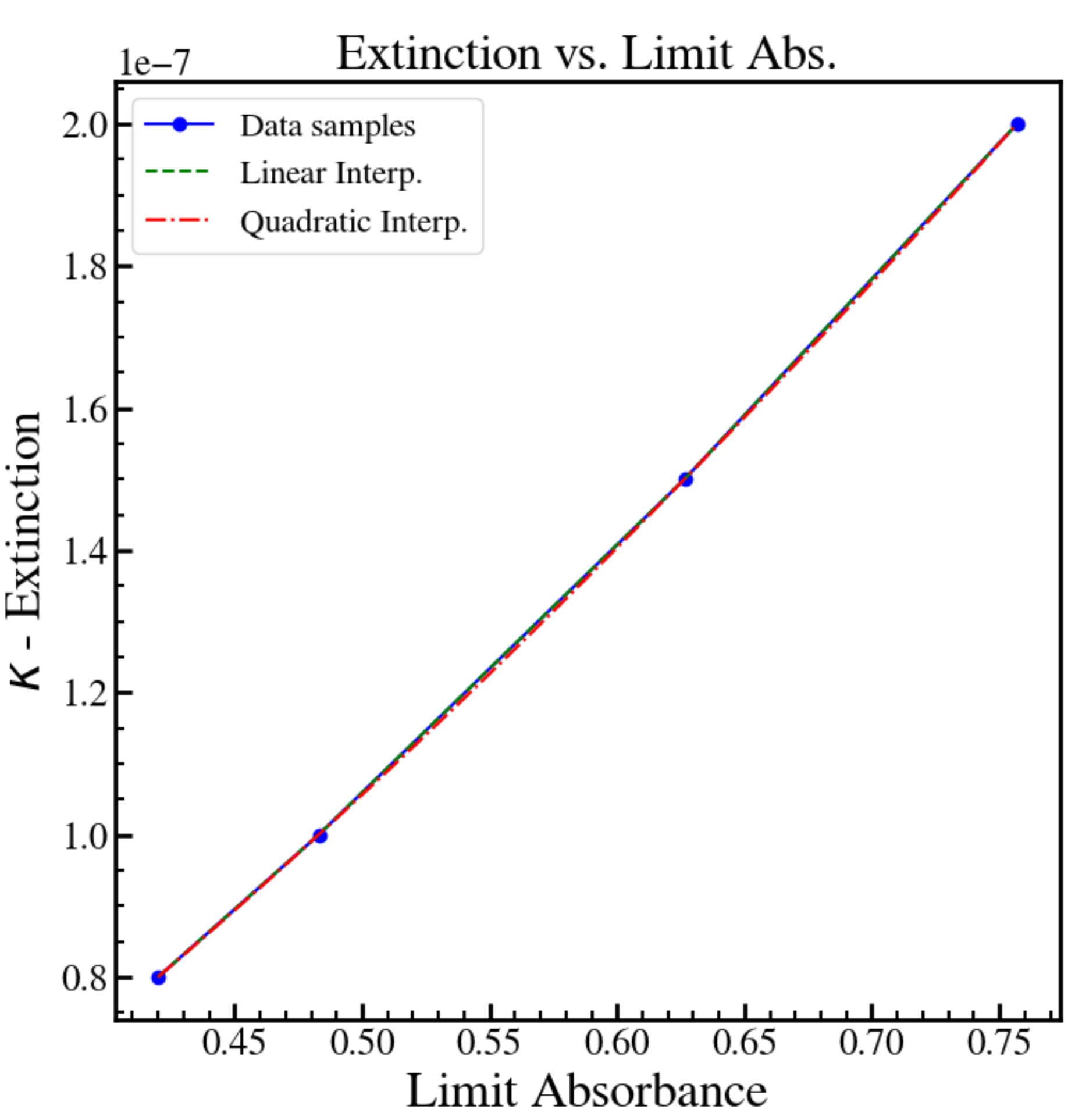"} % Includi l'immagine (sostituisci con il tuo file)
\caption{The doped-Silica extinction  as a function of generalized {\em Carniglia's Limit}, computed by using Fig. \ref{fig:fig5} (same parameters). 
Both linear (green dashed line) and quadratic (red dash-dotted line) interpolations are shown.
This interpolation allows estimating the critical extinction coefficient required to reach a specific target absorbance limit. 
The extrapolated values indicate that achieving the target limit absorbance of $0.5$ ppm requires $\kappa \approx 1.056 \times 10^{-7}$ (using linear interpolation) 
or $\kappa \approx 1.059 \times 10^{-7}$ (using quadratic interpolation). }
\label{fig:fig6}
\end{figure}

\begin{figure}[h] % Opzioni di posizionamento: qui, top, bottom, pagina di floats
\centering % Centra la figura orizzontalmente
\hspace*{-0.5cm}\includegraphics[width=0.47\textwidth]{"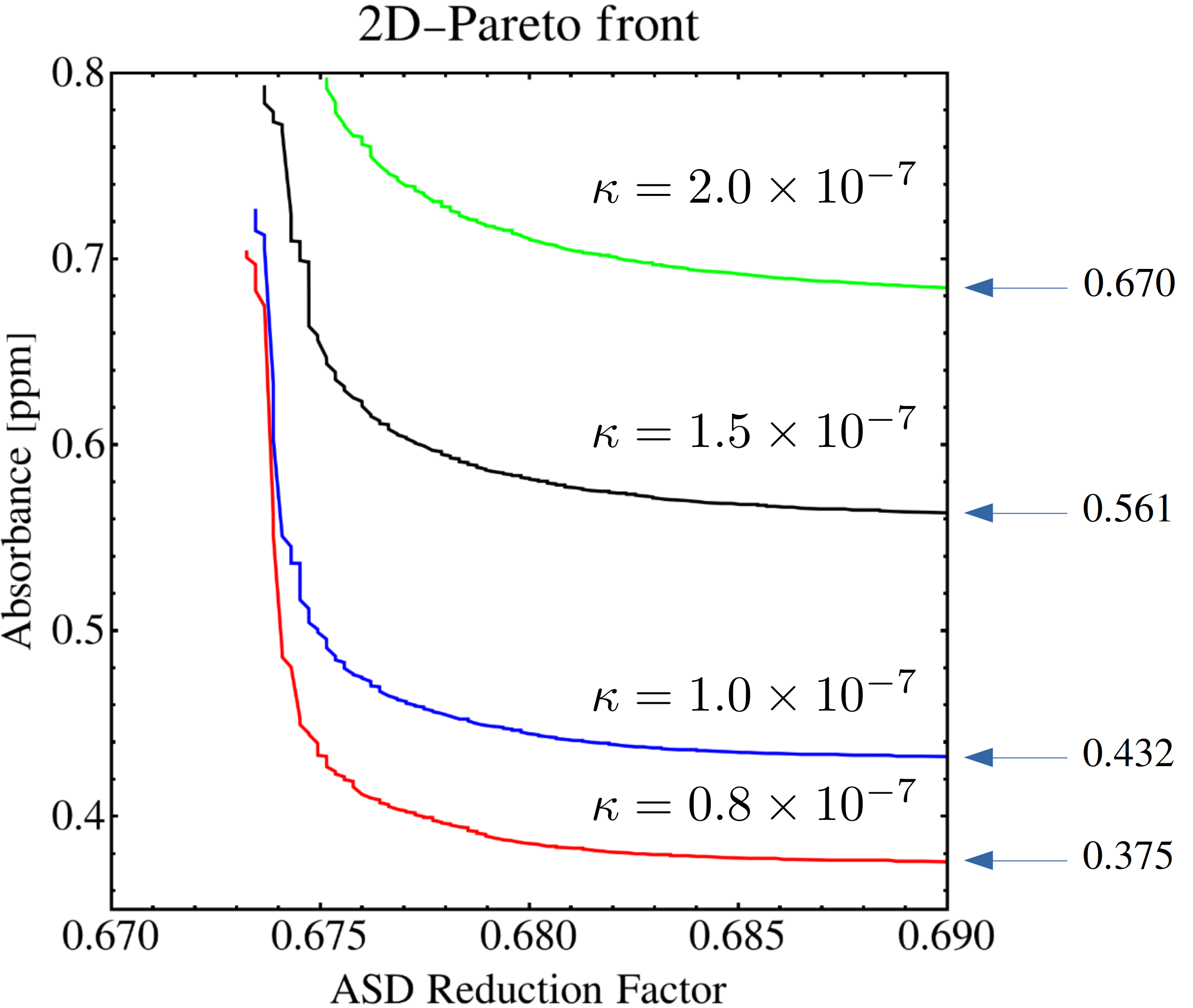"} % Includi l'immagine (sostituisci con il tuo file)
\caption{Pareto front Absorbance versus ASD RF for different values of doped-Silica extinction $\kappa$. The refractive index is $n_r=1.97$ and  
the maximum number of doublets is equal to $40$. The generalized {\em Carniglia's Limit} 
(i.e. the stack limit absorption) is displayed ont the right near the arrows (for each curve).}
\label{fig:fig7}
\end{figure}

\begin{figure}[h] % Opzioni di posizionamento: qui, top, bottom, pagina di floats
\centering % Centra la figura orizzontalmente
\includegraphics[width=0.47\textwidth]{"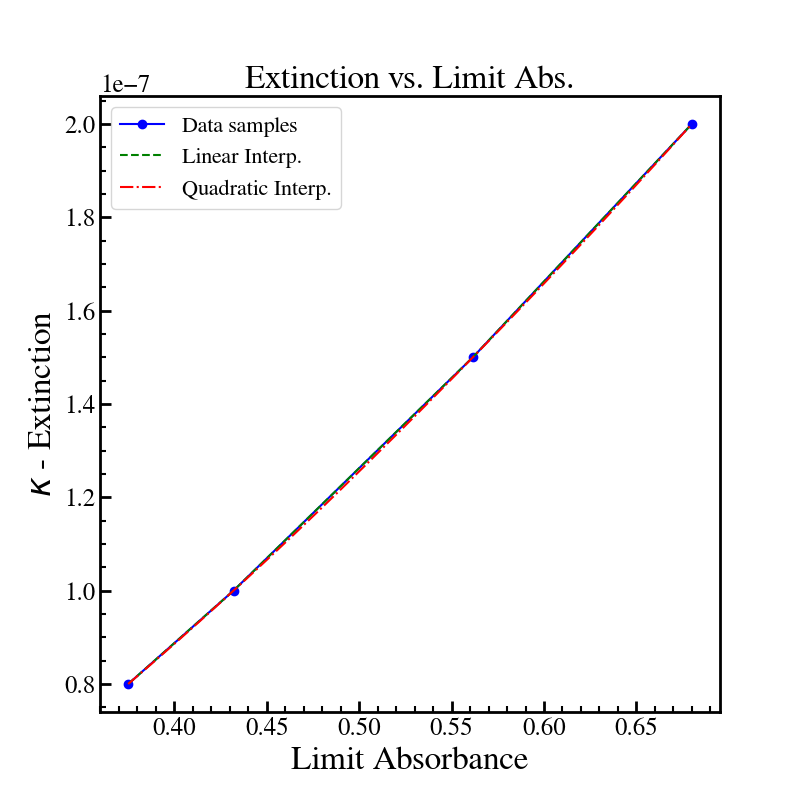"} % Includi l'immagine (sostituisci con il tuo file)
\caption{The doped-Silica extinction  as a function of {\em Carniglia's Limit}, computed by using Fig. \ref{fig:fig7} (same parameters).
Both linear (green dashed line) and quadratic (red dash-dotted line) interpolations are shown.
This interpolation allows estimating the critical extinction coefficient required to reach a specific target absorbance limit. 
The extrapolated values indicate that achieving the target limit absorbance of 0.5 ppm requires $\kappa \approx 1.2556 \times 10^{-7}$ (using linear interpolation) 
or $\kappa \approx 1.2626 \times 10^{-7}$ (using quadratic interpolation).}
\label{fig:fig8}
\end{figure}
The maximum value of the number of doublets was set to $40$.
In the Fig.  \ref{fig:fig5}, on the right is written the value of the generalized CL  for each curve, this value as a function of doped silica extinction is shown in Fig. \ref{fig:fig6}.
As shown in Figure \ref{fig:fig6}, interpolation—either linear or quadratic—of the simulation data allows us to estimate the critical extinction coefficient of doped silica below which the target absorbance limit of $0.5$ ppm
 becomes attainable for the considered coating structure. This analysis provides a practical criterion (though approximate) in order to meet ultra-low absorption requirements in advanced optical coatings.
Interpolation of the data in Fig.  \ref{fig:fig6}  gives the result for the limit extinction of doped silica of $1.06 \times 10^{-7}$. 
The value is of the same order as that estimated by analyzing the KL for the binary Silica /Silica doped coating.

Figures \ref{fig:fig7} and \ref{fig:fig8} report the same extinction analysis discussed previously, now applied to a more optimistic scenario in which the doped silica layer is assumed to have a refractive index of $n_r=1.97$. 
While this higher index slightly modifies the system's optical response, the overall result remains consistent with earlier findings. In particular, 
the critical extinction coefficient required to achieve the $0.5$ ppm absorbance threshold shifts modestly to approximately $\kappa = 1.26 \times 10^{-7}$,
that is of the same order of previous result.

\begin{figure*}[t] % Opzioni di posizionamento: qui, top, bottom, pagina di floats
\centering % Centra la figura orizzontalmente
\hspace*{-0.5cm}\includegraphics[width=0.9\textwidth]{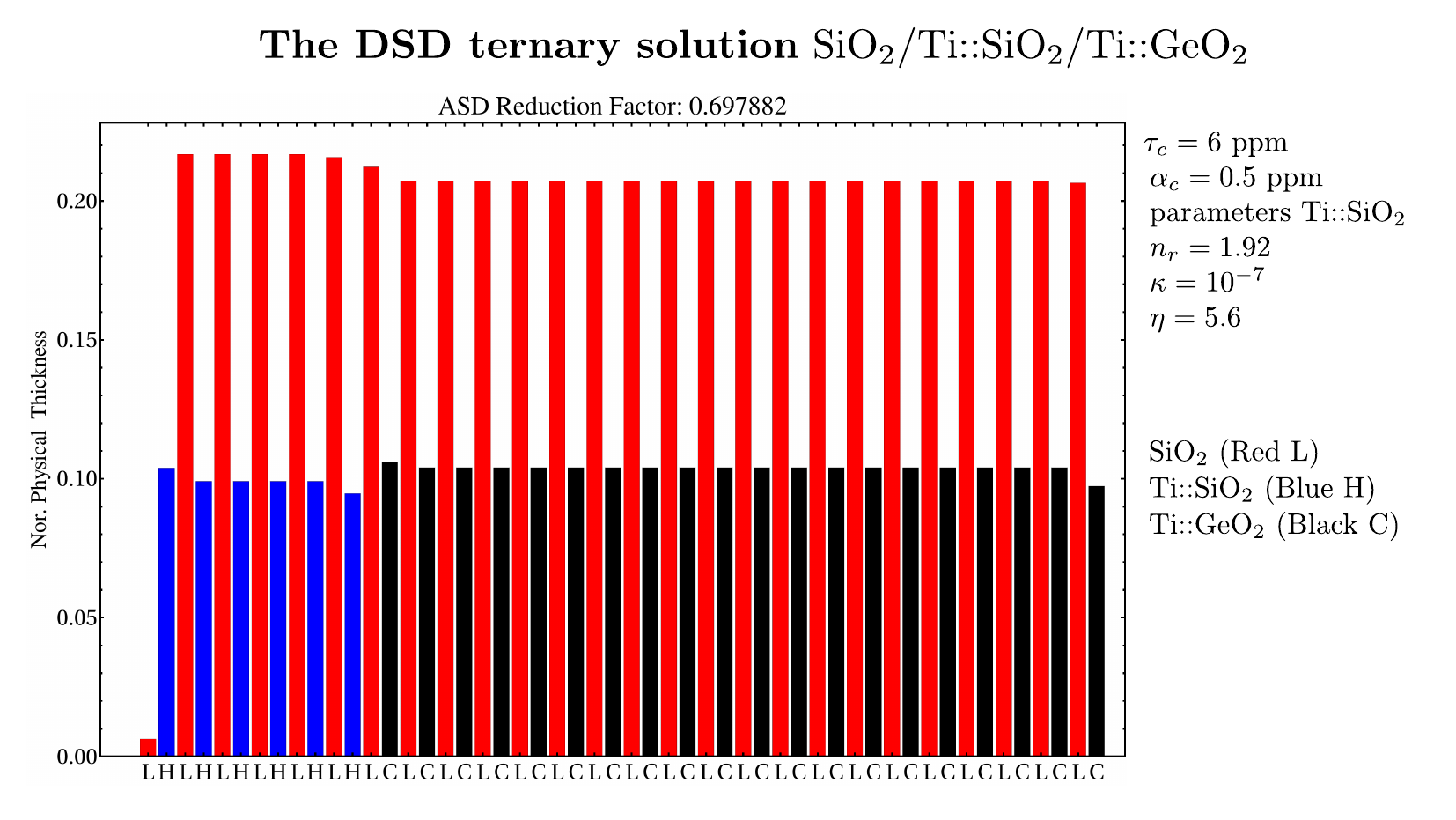} % Includi l'immagine (sostituisci con il tuo file)
\caption{Optimized layer structure for a DSD ternary coating (SiO$_2$/Ti::SiO$_2$/Ti::GeO$_2$). 
Bar heights indicate normalized physical thickness for the layer sequence shown
 (L=SiO$_2$, red; H=Ti::SiO$_2$, blue; C=Ti::GeO$_2$, black). The resulting ASD RF is $0.697882$, 
 calculated using the Ti::SiO$_2$ parameters $n_r=1.92$ and $\kappa=10^{-7}$, and $\eta_{Ti::SiO_2}=5.6$ among others listed.}
\label{fig:fig9}
\end{figure*}
 
%%%%%%%%%%%%%%%%%%%%%%%%%%%%%%%%%%%%%%%%%%%%%%%%%%%%%%%%%%%%%%%%%%%%%%%%%%%%%%%%%%%%%%%%%%%%%%%%%%%%%%%%%%%%%%%%%%%%%%%%%%%%%%%%%%%
\section{Ternary Coating: Silica/Ti::Silica/Ti::GeO$_2$ DSD design.}
%%%%%%%%%%%%%%%%%%%%%%%%%%%%%%%%%%%%%%%%%%%%%%%%%%%%%%%%%%%%%%%%%%%%%%%%%%%%%%%%%%%%%%%%%%%%%%%%%%%%%%%%%%%%%%%%%%%%%%%%%%%%%%%%%%%

From the analyses presented in the previous sections, we derive two key reference parameters that will serve as the starting point for the fine optimization of the ternary coating structure. 
First, the effective Braginsky coefficient associated with the doped silica layer, denoted by $\eta_{Ti::SiO_2}$, is found to be reasonably set at $5.6$.

Secondly, to be viable in a ternary coating architecture, particularly one targeting ultra-low absorbance levels, doped silica requires an extinction coefficient ($\kappa$) on the order of $10^{-7}$, 
assuming a real part of its refractive index ($n_r$) of $1.92$.
This extremely low absorption value represents a critical threshold below which the doped silica can meaningfully participate in the structure without compromising the overall absorbance target (i.e. $0.5$ ppm).

These two parameters provide a realistic and application-driven starting point for the fine-scale optimization of ternary designs capable of achieving sub-ppm absorption performance.

Layer structure of an optimized ternary coating using the DSD design, composed of Silica, Titania-doped Silica, and Titania-doped GeO$_2$ is shown in Fig. \ref{fig:fig9}.
To obtain the design, we imposed constraints on transmissivity ($\tau_c \le 6$ ppm) and absorbance ($\alpha_c \le 0.5$ ppm).
The vertical axis represents the normalized physical thickness of each layer (normalization w.r.t. $\lambda_0=1064$ nm), 
arranged in the sequence shown on the horizontal axis. 
The layer sequence shown is deposited onto the substrate, which is situated on the right side. This specific design achieves the ASD RF  $0.698$.
The design parameters shown, including refractive index ($n_r=1.92$) and extinction coefficient ($\kappa=10^{-7}$) for Ti::SiO$_2$, were used in the optimization process, performed by our algorithm \cite{BORGVP, VPbinary}.

\begin{figure}[h] % Opzioni di posizionamento: qui, top, bottom, pagina di floats
\centering % Centra la figura orizzontalmente
\hspace*{-0.5cm} \includegraphics[width=0.6\textwidth, trim={1.5cm 0cm 0cm 0cm},clip]{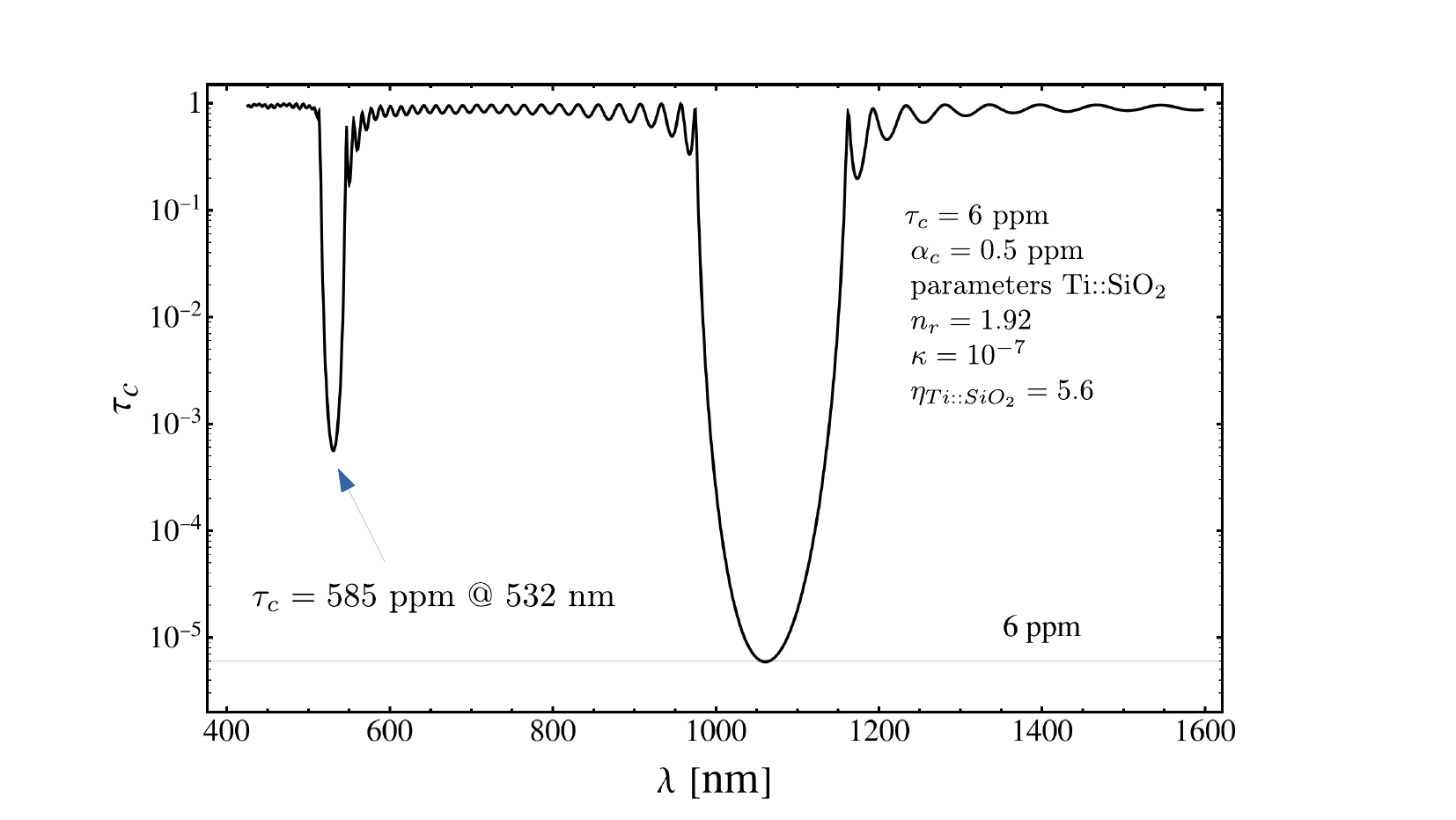} % Includi l'immagine (sostituisci con il tuo file)
\caption{Computed transmittance spectrum ($\tau_c$ coating transmittance) versus wavelength ($\lambda$ nm) for the optimized DSD ternary coating (SiO$_2$/Ti::SiO$_2$/Ti::GeO$_2$). 
The plot shows high-reflectivity bands (low transmittance minima) near $532$ nm and $1064$ nm (primary band).
 The calculated transmittance at $532$ nm is indicated as $585$ ppm. The design parameters used for Ti::SiO$_2$ ($n_r=1.92, \kappa=10^{-7}$) and target constraints (on $\tau_c$, and $\alpha_c$) are listed. 
 The horizontal dashed line represents a level of 6 ppm (project target).}
\label{fig:fig10}
\end{figure}

\begin{figure}[h] % Opzioni di posizionamento: qui, top, bottom, pagina di floats
%\centering % Centra la figura orizzontalmente
\hspace*{-1.2cm}\includegraphics[width=0.65\textwidth]{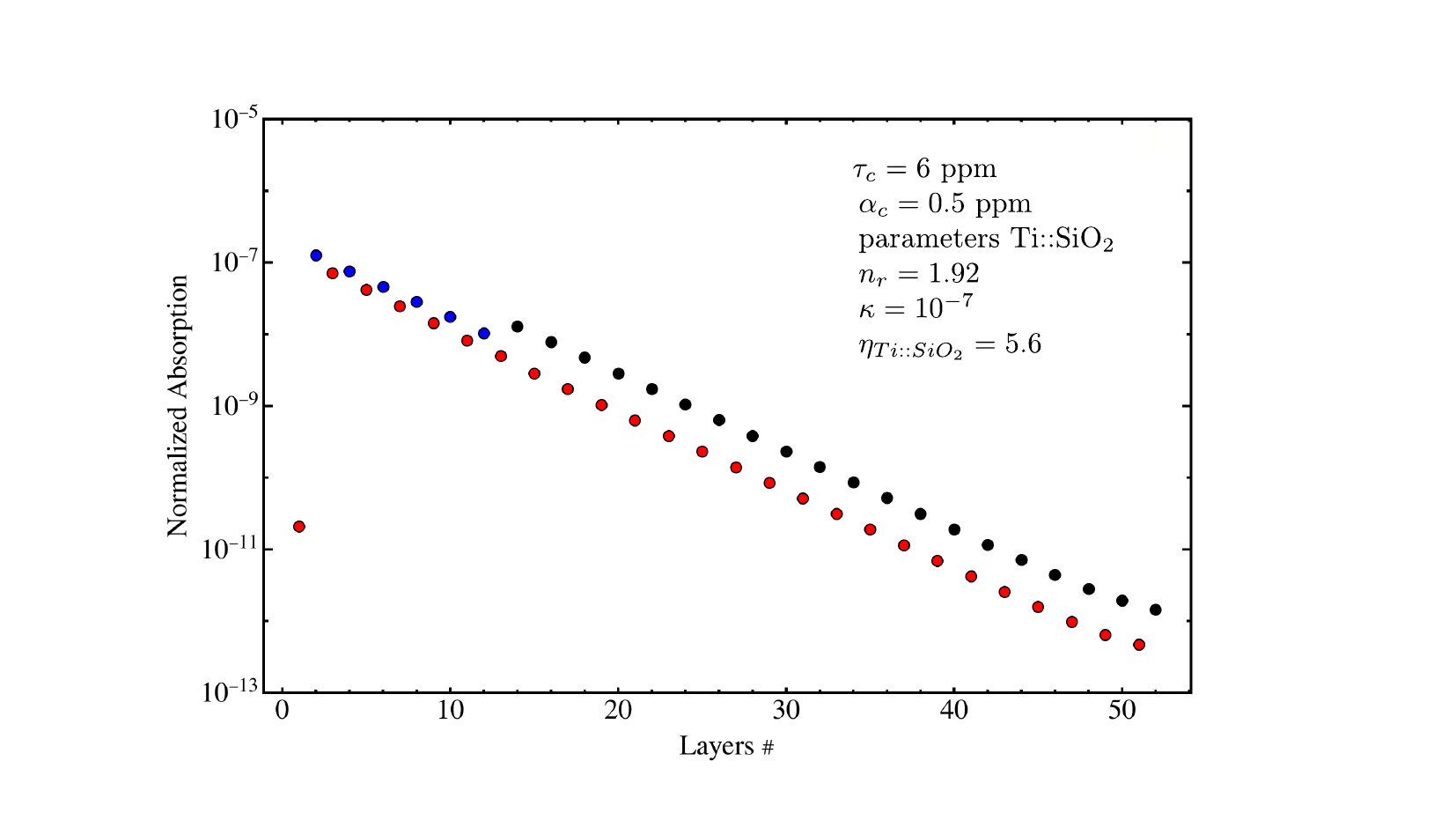} % Includi l'immagine (sostituisci con il tuo file)
\caption{Normalized absorption profile within the DSD ternary coating structure versus layer number (starting from the incident medium on the left side ). 
The plot shows the relative absorption contribution per layer. Different colors likely represent the different materials: SiO$_2$ (red dots), Ti::SiO$_2$ (blue dots), and Ti::GeO$_2$ (black dots), 
corresponding to the layer types in the structure. Simulation parameters are the same of Fig. \ref{fig:fig10}.
}
\label{fig:fig11}
\end{figure}

The optical performance and internal absorption distribution of the optimized DSD ternary coating design are presented in Figs. \ref{fig:fig10} and \ref{fig:fig11} respectively. 
Figure \ref{fig:fig10} shows the calculated transmittance spectrum (assuming that no dispersion affects refractive indices).
The plot shows high-reflectivity bands (low transmittance minima) developed near the interferometer's primary operating wavelength of $1064$ nm
and a secondary wavelength of $532$ nm (half-wavelength band). The optimization procedure was explicitly 'single-band', targeting only the performance requirements at $1064$ nm.
The resulting calculated transmittance at $532$ nm is 585 ppm, as indicated. The design parameters used for Ti::SiO$_2$ ($n_r=1.92, \kappa=10^{-7}$, $\eta_{Ti::SiO_2}=5.6$) and target constraints ($\tau_c$, $\alpha_c$) are listed. 
The choice for a single-band optimization strategy, rather than attempting to meet requirements at both wavelengths simultaneously (dual-band), is crucial. 
Imposing additional constraints during the optimization process specifically to increase the transmittance at $532$ nm would likely worsen the achieved performance 
(e.g., increase losses or reduce reflectivity) in the primary, and more critical, $1064$ nm band.

Figure \ref{fig:fig11} provides insight into the internal loss mechanisms of the coating by showing the distribution of absorption throughout the stack..
The {\em Normalized Absorption} plotted for each layer represents the ratio of power absorbed in that specific layer to the total power incident on the coating. 
Crucially, these normalized values allow for the direct calculation of the absolute power delivered as heat within each layer simply by multiplying them by the known incident laser power. 
This detailed knowledge of power dissipation per layer is fundamental for subsequent thermal analysis \cite{Hello}. 
Such analysis aims to predict the temperature distribution and overall temperature increase within the coating when subjected to high incident optical power, which is vital for assessing performance limitations like thermal lensing, 
mechanical stress, or potential laser-induced damage \cite{EMdamage}. As expected, the absorption is higher near the front surface (low layer numbers) where the field intensity is greatest within the high-reflectivity structure.

\begin{figure}[h] % Opzioni di posizionamento: qui, top, bottom, pagina di floats
%\centering % Centra la figura orizzontalmente
\includegraphics[width=0.47\textwidth]{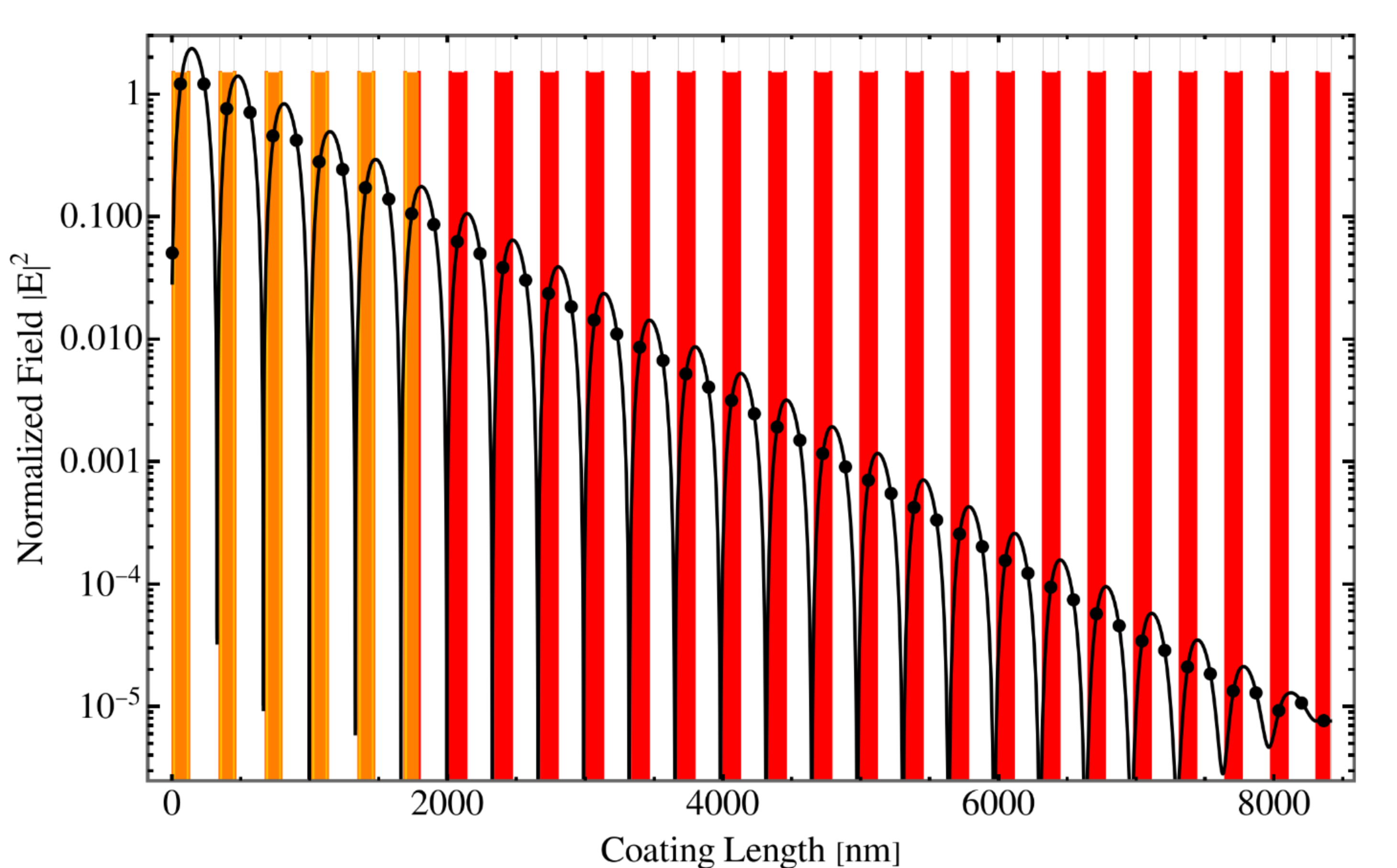} % Includi l'immagine (sostituisci con il tuo file)
\caption{Profile of the normalized squared electric field intensity  ($|E|^{2}/|E_{inc}|^{2}$) within the optimized DSD ternary coating structure,
plotted against the physical distance from the coating's front surface (Coating Length [nm]) at the design wavelength. 
The continuous black line shows the calculated field intensity profile. The vertical colored bars represent the individual layers according to the legend: Ti::GeO$_2$ (red), SiO$_2$ (white/no bar shown, but implied by spacing), and Ti::SiO$_2$ (orange). 
The black circles indicate the average normalized field intensity calculated within each corresponding layer.}
\label{fig:fig12}
\end{figure}

\begin{figure*}[t] % Opzioni di posizionamento: qui, top, bottom, pagina di floats
\centering % Centra la figura orizzontalmente
\hspace*{-0.5cm}\includegraphics[width=1\textwidth]{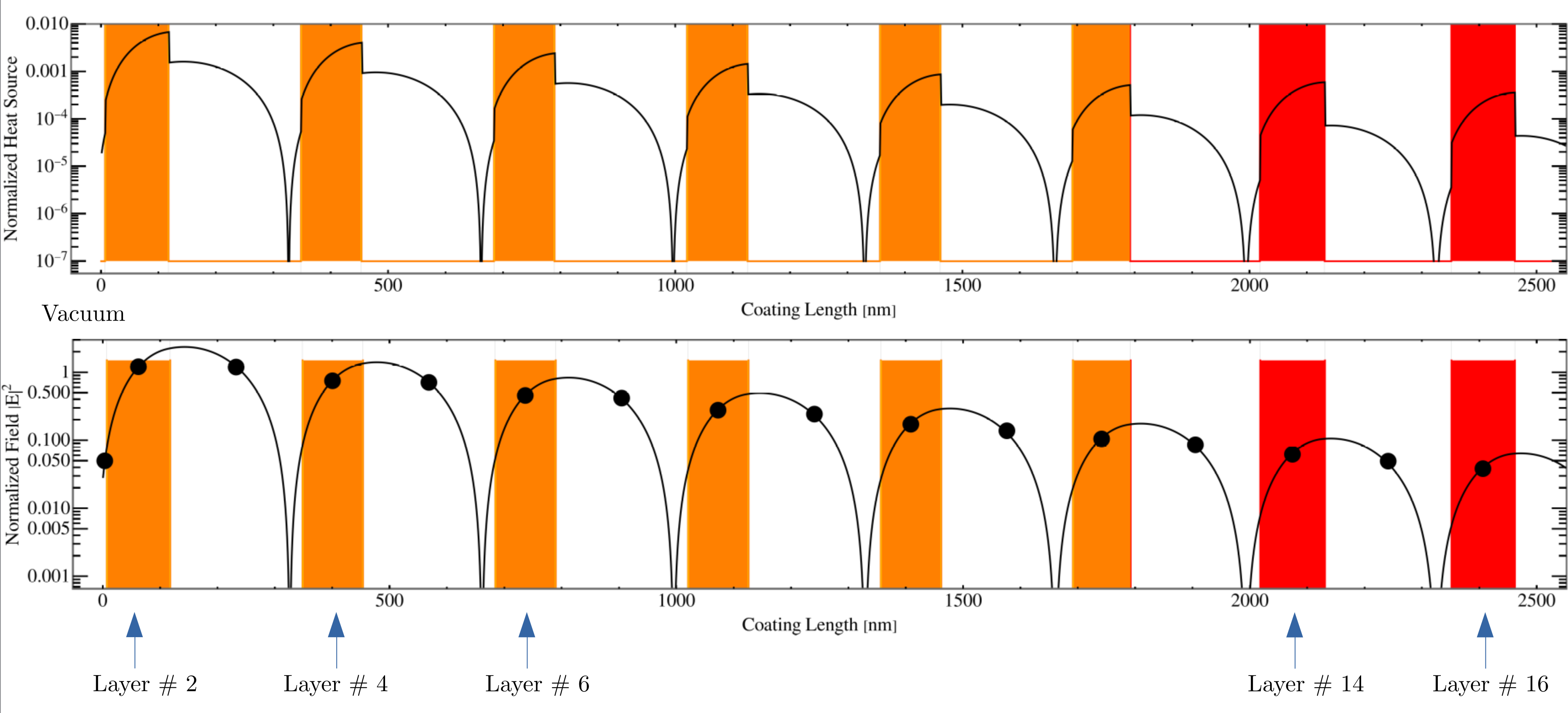} % Includi l'immagine (sostituisci con il tuo file)
\caption{Closeup of the normalized volumetric heat source profile (top panel) and the corresponding normalized electric field intensity squared profile (bottom panel)
 within the first $\sim 2500$ nm of the DSD ternary coating structure. 
 The abscissa represents the coating length [nm]. \textbf{Top Panel:} Shows the Normalized Heat Source ($\dot{Q}/|E_{inc}|^2$), computed using the local field 
 intensity and material properties ($\kappa$, $n_r$) as per the formula provided in the text. 
\textbf{Bottom Panel:} Closeup the Normalized Field intensity ($|E|^2/|E_{inc}|^2$) profile from Fig. \ref{fig:fig12}, on the initial layers, with average field values per layer 
indicated by black circles and specific layer numbers marked. The color of bar indicate the same materials of Fig. \ref{fig:fig12}.}
\label{fig:fig13}
\end{figure*}

To provide a deeper understanding of why the power absorption shown in Figure \ref{fig:fig11} is distributed in that specific manner, 
Figure \ref{fig:fig12} presents the profile of the normalized electric field intensity squared ($|E|^{2}/|E_{inc}|^{2}$) throughout the DSD ternary coating structure at the design wavelength ($\lambda_0=1064$ nm). 
This quantity is crucial because the local power absorption within any given layer is related to this field intensity and the material's extinction coefficient (imaginary part of refractive index). 
This is computed by the formula $\displaystyle \dot{Q}= \frac{2  \pi c}{\lambda_0} \epsilon_0 \kappa n_r | E|^2$, where $c$ is the speed of light and $\epsilon_0$ the vacuum electric permeability.
The extinction coefficient $\kappa$  and refractive index $n_r$ vary according to the specific material used for each layer.
The absorption heat source is clearly discontinuous with depth due to the layered nature of the structure, as illustrated in the detailed view of Fig. \ref{fig:fig13}.

Figures \ref{fig:fig12} and \ref{fig:fig13}  therefore visualizes the spatial distribution of the electromagnetic energy density within the coating.
 Both clearly shows the standing wave pattern formed by the interference of incident and reflected waves, and importantly, it reveals the rapid decay of the field intensity as the light penetrates into the high-reflectivity stack.
 The regions with high field intensity peaks are where most of the absorption will occur if the material in that location has a non-negligible extinction coefficient. 
 The average field intensity within each layer, also indicated, provides a useful metric that relates directly to the total absorption per layer shown previously in Fig. \ref{fig:fig11}.

\begin{figure}[h] % Opzioni di posizionamento: qui, top, bottom, pagina di floats
\centering % Centra la figura orizzontalmente
\includegraphics[width=0.395\textwidth]{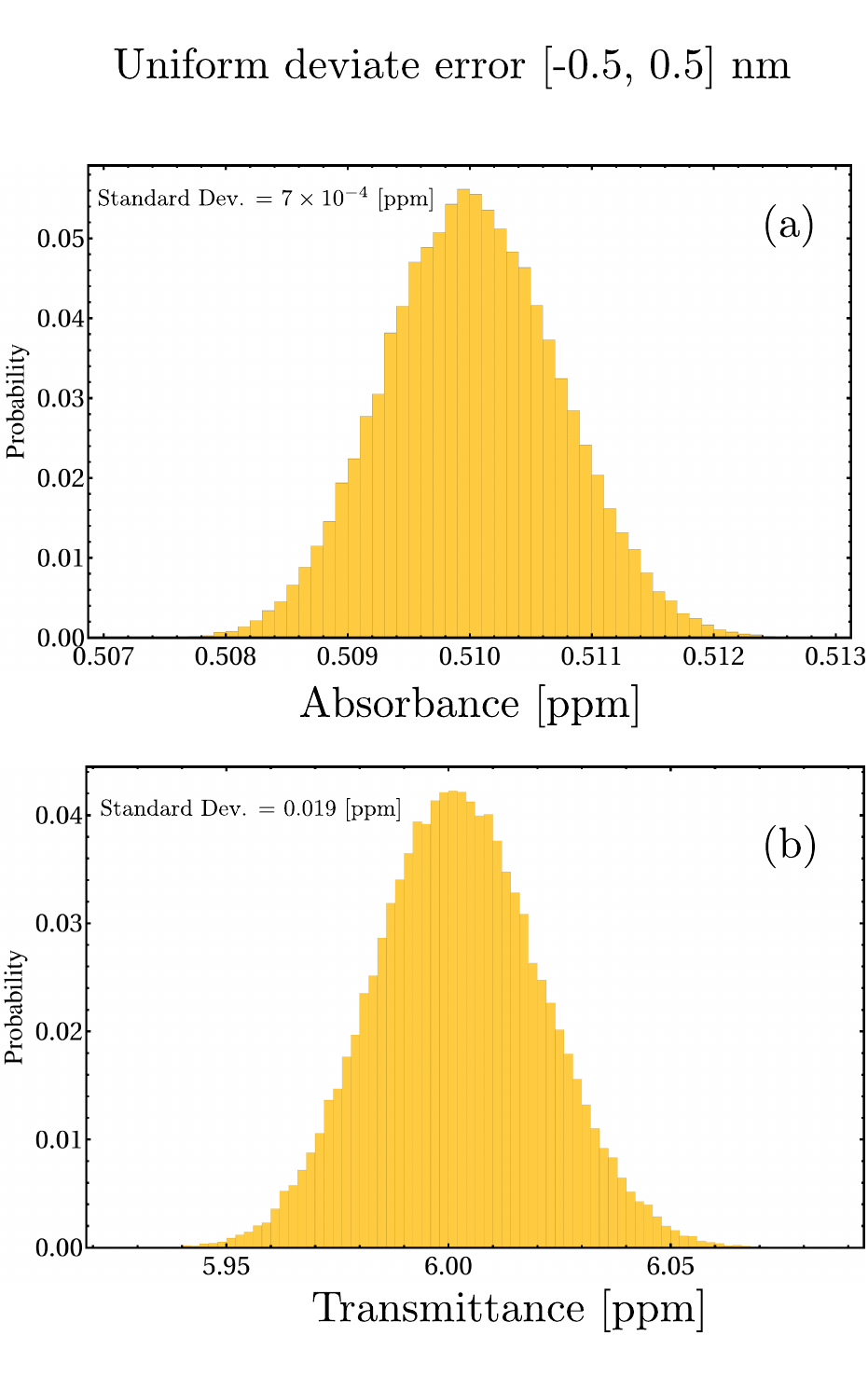} % Includi l'immagine (sostituisci con il tuo file)
\caption{The statistical frequency distributions of absorbance (panel a) and transmittance (panel b), expressed in parts per million, ppm, of proposed DSD ternary design.
The distributions are obtained via Monte Carlo simulation with \(10^5\) trials. 
The input parameters assume random variations in the layer thicknesses, uniformly distributed within the interval \([-0.5, 0.5]\) nm around their nominal values.
In the same condition, the standard deviation for ASD RF is $1.29919 \times 10^{-4}$}
\label{fig:fig13bis}
\end{figure}

As a final step, we investigate the robustness of the proposed ternary design to fabrication uncertainties in layer thicknesses.
In Fig. \ref{fig:fig13bis}, we show the statistical frequency distributions of absorbance (panel a) and transmittance (panel b), expressed in parts per million (ppm),
 for the proposed ternary design. 
The distributions are obtained via Monte Carlo simulation with \(10^5\) trials. The input parameters assume random variations in the layer thicknesses, 
uniformly distributed within the interval \([-0.5, 0.5]\) nm around their nominal values.

The resulting curves illustrate the optical coating’s sensitivity to microscopic thickness fluctuations, showing a strong concentration of probability around the expected mean values. 
The distributions appear approximately Gaussian, indicating a robust optical response to random layer thickness variations within this specified tolerance range. 
Under these conditions, the standard deviation of the estimated ASD RF is calculated to be  $1.29919 \times 10^{-4}$.

\begin{figure}[h] % Opzioni di posizionamento: qui, top, bottom, pagina di floats
\centering % Centra la figura orizzontalmente
\includegraphics[width=0.40\textwidth]{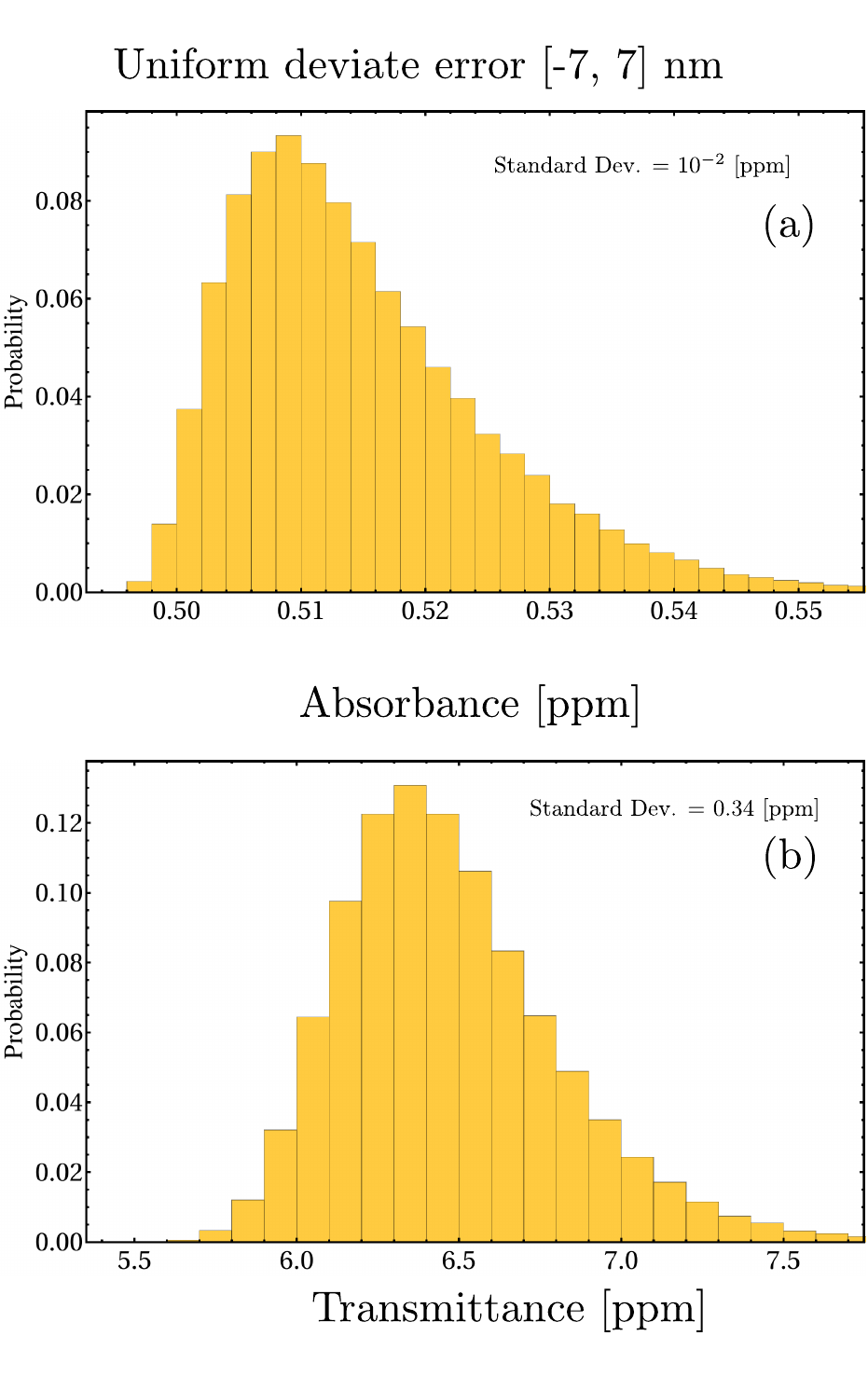} % Includi l'immagine (sostituisci con il tuo file)
\caption{The statistical frequency distributions of absorbance (panel a) and transmittance (panel b), expressed in parts per million, ppm, of proposed DSD ternary design.
The distributions are obtained via Monte Carlo simulation with \(10^5\) trials. 
The input parameters assume random variations in the layer thicknesses, uniformly distributed within the interval \([-7, 7]\) nm around their nominal values.
In the same condition, the standard deviation for ASD RF is $1.81116 \times 10^{-3}$}
\label{fig:fig14bis}
\end{figure}

Figure \ref{fig:fig14bis} is generated under the same simulation conditions as Fig. \ref{fig:fig13bis}, except that the range of uncertainty variation is significantly wider: ([-7, 7]) nm 
around the nominal layer thicknesses.  
Compared to the case with smaller uncertainty, the distributions shown here deviate significantly from a Gaussian profile, indicating that the system has entered a regime in which the optical 
response becomes nonlinear with respect to thickness variations. The standard deviations of absorbance and transmittance, particularly the latter, are significantly larger, 
underscoring the increased sensitivity of the coating’s optical properties under these larger fabrication tolerances. 
The standard deviation for the ASD RF in this case increased substantially to $1.81116 \times 10^{-3}$.

The estimated standard deviations for the ASD RF provide valuable insight into the practical significance of design improvements. The relatively small standard deviation obtained under the tighter tolerance 
($1.29919 \times 10^{-4}$) suggests that a reduction in the ASD RF value, which is appreciable down to the third decimal place, would not be obscured by the level of fabrication imprecision considered. 
This confirms that targeted improvements in the design's performance at this level are meaningful and robust against expected manufacturing variability within the tighter tolerance range. 
The higher manufacturing uncertainty results in a higher standard deviation ($1.81116 \times 10^{-3}$), which may mask the finer improvements, although improvements to the second decimal place are still appreciable.

%%%%%%%%%%%%%%%%%%%%%%%%%%%%%%%%%%%%%%%%%%%%%%%%%%%%%%%%%%%%%%%%%%%%%%%%%%%%%%%%%%%%%%%%%%%%%%%
\section{Alternative Strategy: relaxing absorbance constraint}
%%%%%%%%%%%%%%%%%%%%%%%%%%%%%%%%%%%%%%%%%%%%%%%%%%%%%%%%%%%%%%%%%%%%%%%%%%%%%%%%%%%%%%%%%%%%%%%s
Our primary design goal for the mirror coatings is to achieve an ASD RF of $0.5$, 
aiming to minimize the contribution of coating thermal noise in future gravitational wave detectors. 
However, as demonstrated by the results presented earlier (e.g., Fig. \ref{fig:fig9}), the optimization aimed at finding the best DSD structure under the very tight $0.5$ ppm
absorbance constraint yielded a best achievable ASD RF of approximately $0.698$.

While this represents a substantial reduction compared to standard coatings, indicating a competitive potential with the experimental results obtained for Ti::GeO$_2$ coatings (ASD RF $\sim 0.74$ \cite{TiGeO}) 
in the absence of problems introduced during deposition (such as material incompatibility, defects, or fractures), 
the result unfortunately still falls short of the required $0.5$ target for the advanced LIGO and Virgo project.

This outcome indicates that achieving the desired ASD RF is not feasible with the current ternary system and the DSD approach when simultaneously enforcing such a low absorbance limit. 
Consequently, a change in strategy is required. We will now investigate the potential improvements by relaxing the constraint on absorption, 
specifically by increasing the target absorbance value to $\alpha_c \le 1$ ppm in subsequent optimizations, potentially still based on the DSD approach.

\begin{figure*}[t] % Opzioni di posizionamento: qui, top, bottom, pagina di floats
\centering % Centra la figura orizzontalmente
\includegraphics[width=0.9\textwidth]{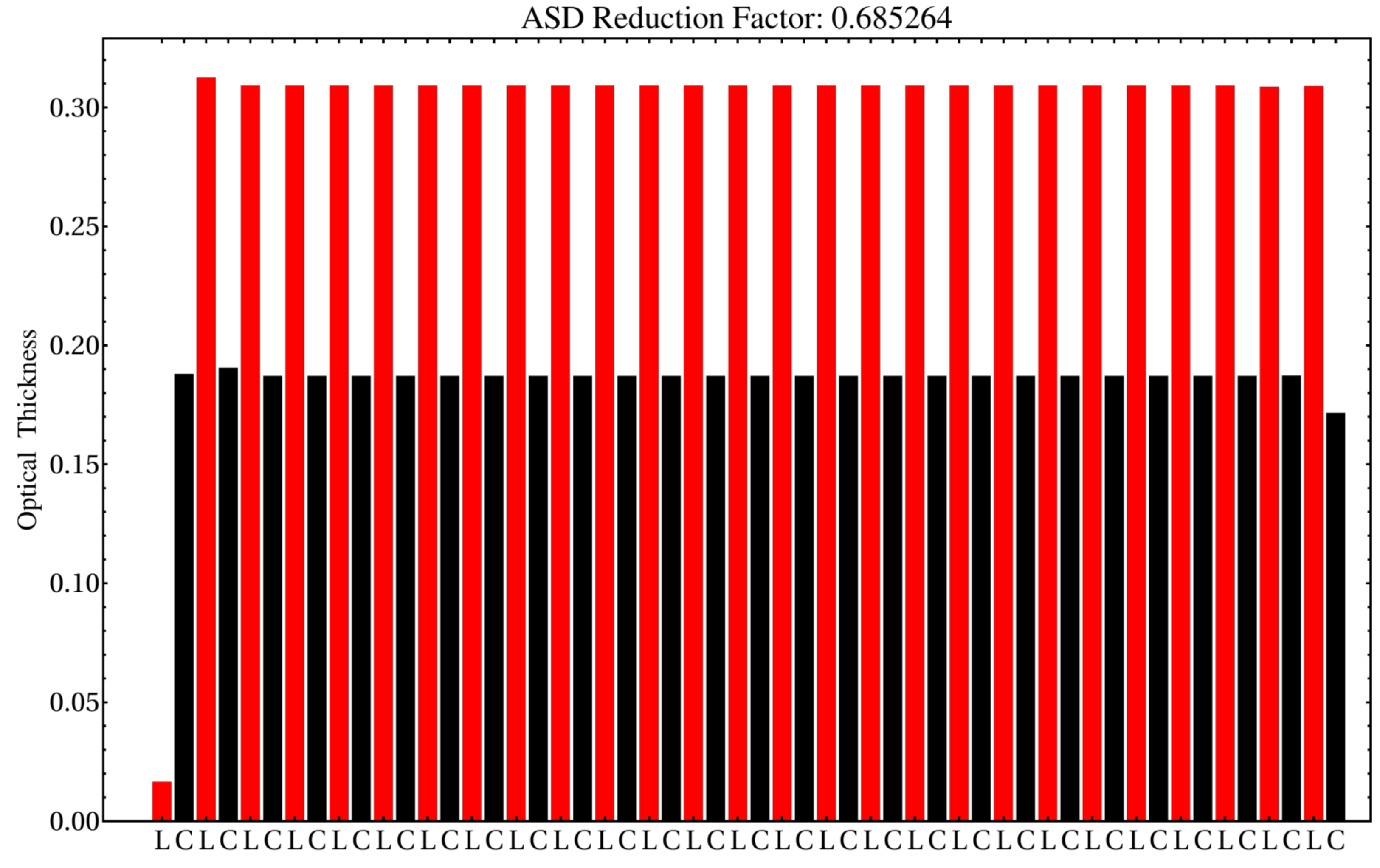} % Includi l'immagine (sostituisci con il tuo file)
\caption{Layer structure (Optical Thickness vs. Layer sequence L=SiO$_2$, C=Ti::GeO$_2$) for a binary coating identified during optimization under a relaxed absorbance constraint of $\alpha_c = 1$ ppm. 
This soptimal solution achieves an ASD Reduction Factor of $0.685264$ with a computed total absorbance of approximately $0.8$ ppm. The color code is L=Red, C=Black. }
\label{fig:fig14}
\end{figure*}

\begin{figure*}[t] % Opzioni di posizionamento: qui, top, bottom, pagina di floats
\centering % Centra la figura orizzontalmente
\includegraphics[width=0.9\textwidth]{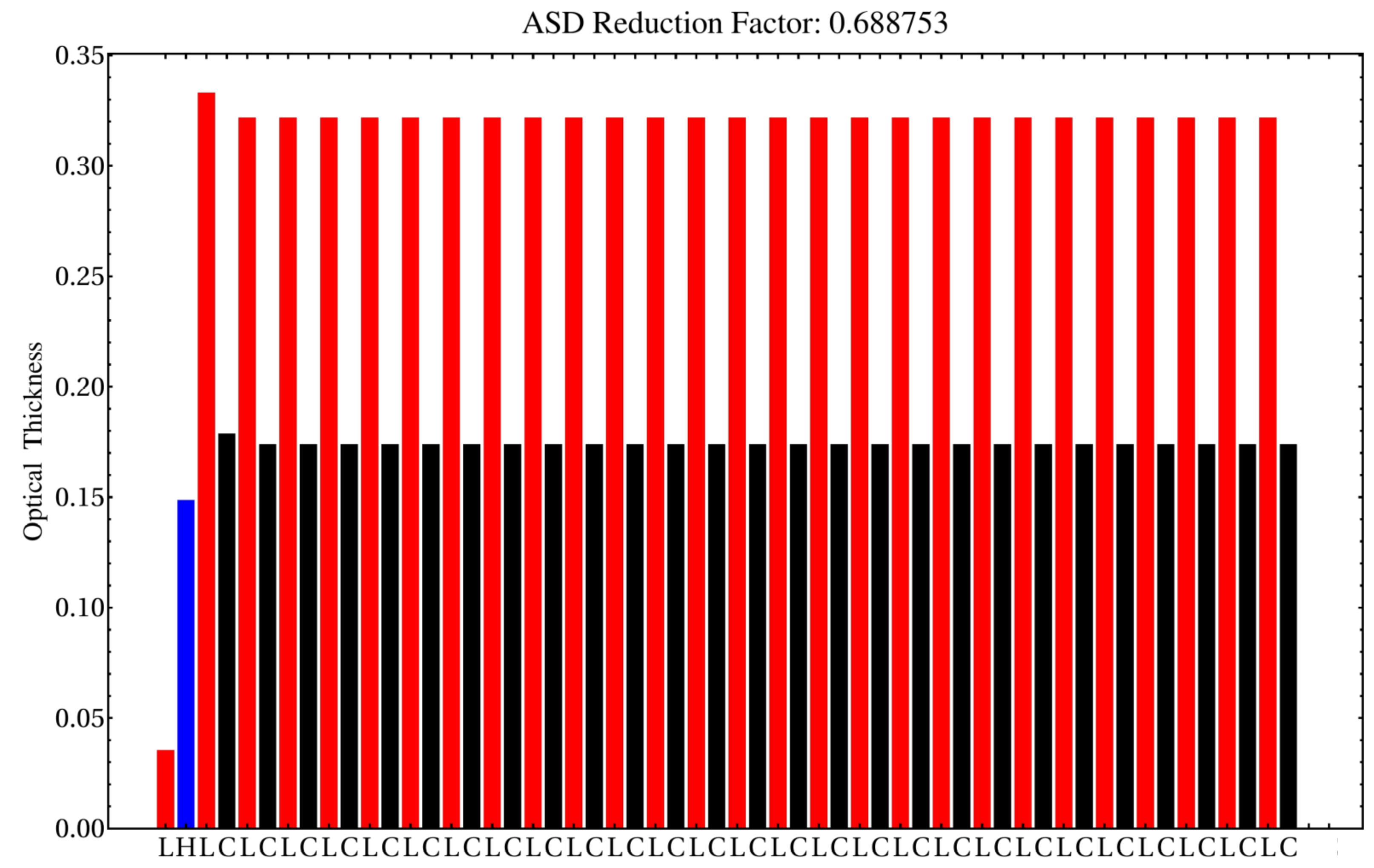} % Includi l'immagine (sostituisci con il tuo file)
\caption{Layer structure (Optical Thickness vs. Layer sequence L=SiO$_2$,H=Ti::SiO$_2$, C=Ti::GeO$_2$) for the sub-optimal DSD coating identified by the optimization algorithm under a 
relaxed absorbance constraint of $\alpha_c = 1$ ppm. This solution yields the best ASD Reduction Factor of 0.688753. Its calculated total absorbance is approximately 0.9 ppm, remaining below the 1 ppm constraint.
The color code is L=Red, H=Blue, and C=Black.}
\label{fig:fig15}
\end{figure*}

The optimization was re-run with the same material parameters (for SiO$_2$, Ti::SiO$_2$, and Ti::GeO$_2$) but setting the target absorbance limit higher, at $\alpha_c \le 1$ ppm. 
Figures \ref{fig:fig14} and \ref{fig:fig15} present key results from this revised optimization.
Figure \ref{fig:fig14} shows the structure identified by the optimization algorithm as yielding the absolute lowest (best) ASD RF, achieving a value of $0.688753$.
 A notable outcome under this relaxed constraint is that the optimal strategy involves only two materials: the low-index SiO$_2$ (red bars) and the higher-loss, high-index Ti::GeO$_2$ (black bars). 
 The intermediate Ti::SiO$_2$ material is completely omitted in this optimal design. It is also important to note that this solution does not saturate the imposed constraint; 
 its calculated absorbance reaches approximately $\alpha_c \sim 0.9$ ppm,  remaining below the $1$ ppm limit.
Figure \ref{fig:fig15}   presents a different, slightly sub-optimal solution found during the same optimization run. This structure yields a slightly higher (worse) ASD RF of $0.685264$. 
Although sub-optimal in terms of ASD RF, this solution is interesting because it achieves its performance with a lower total absorbance, calculated to be around $\alpha_c \sim 0.8$ ppm. 
In this case, the solution uses all materials (SiO$_2$, Ti::SiO$_2$, and Ti::GeO$_2$), though there is only one doublet containing doped silica (Ti::SiO2).
%These results under the relaxed constraint reveal that the optimal DSD structure simplifies to a binary system and highlight a potential trade-off between achieving 
%the absolute minimum ASD RF and maintaining the lowest possible coating absorbance within the allowed limit.

In summary, our optimization studies using the ternary material system (SiO$_2$, Ti::SiO$_2$, Ti::GeO$_2$) within the DSD framework demonstrate a clear limitation.
 Initially, imposing a stringent absorbance constraint of $\alpha_c = 0.5$ ppm resulted in a best achievable ASD Reduction Factor of approximately $0.698$. 
 Subsequently, relaxing the absorbance constraint to $\alpha_c = 1$ ppm, while leading the optimization to favor a binary SiO$_2$/Ti::GeO$_2$ structure, only yielded a marginally improved best ASD RF of approximately $0.689$.
 
Crucially, neither optimization scenario allowed us to reach the target design goal of ASD RF = 0.5. Even when permitting absorbance levels as high as 1 ppm, the fundamental properties of this specific material combination, 
coupled with the DSD design approach, appear insufficient to simultaneously achieve the desired low coating thermal noise (represented by ASD RF = 0.5) and the imposed absorbance limits. 
This strongly suggests that reaching the ambitious 0.5 target will require exploring alternative strategies, such as investigating other materials.

Finally, it is worth noting that recent measurements of both mechanical and optical properties for doped silica \cite{GraemePri} indicate the possibility of depositing this material with 
lower Braginsky coefficient and extinction coefficient comparable to doped germania. 
If confirmed in the actual fabrication of stacks, these results could make the use of doped germania unnecessary, because the binary coatings consisting of silica and doped silica 
with such high quality would exhibit the same performance as ternary coatings made of silica, Ti:GeO$_2$, and doped silica with the parameters given in Table \ref{tab:braginsky_coeff}.
In any case, the results would be no better than those shown in the next paragraph
%% rimuovi appendice B

\section{Alternative Strategy: denser materials}

Our optimization studies using the  DSD approach with the SiO$_2$ / Ti::SiO$_2$ / Ti::GeO$_2$ ternary system have established the current 
performance benchmark achievable with this specific material combination.
 As shown above, the best attained ASD Reduction Factor (ASD RF) reached approximately $0.69$, a significant result representing a 
substantial improvement over conventional designs (if experimentally confirmed).
However, while representing the state-of-the-art for this particular material set, this optimal value still does not meet the ambitious final design goal of ASD RF = 0.5, required for further minimizing thermal noise. 
This suggests that although the DSD approach with these materials offers tangible benefits, pushing beyond the current performance limit towards the 0.5 target necessitates exploring alternative strategies, particularly regarding the materials employed.
 
Therefore, the subsequent phase of our work focuses on assessing the potential of ternary DSD designs that incorporate higher-index materials. 
The strategy remains to combine a low-index material, typically silica (SiO$_2$), with two distinct high-index materials (H$_1$ and H$_2$) chosen specifically for their large refractive indices. 
The rationale is that increasing the refractive index contrast within the coating stack generally offers greater design flexibility.
 This increased flexibility might allow the DSD optimization to find layer thickness configurations that more effectively suppress coating thermal noise, potentially enabling the achievement of the elusive ASD RF =0 .5 goal, while still managing total absorbance within acceptable limits.

In the following we set material H$_1$ = Ti::GeO$_2$ and explore two alternative for H$_2$, the SiN$_x$ (silicon nitride $n_r \sim 2.05$) and aSi (amorphous silicon $n_r \sim 3.2$).

\begin{figure}[h] 
\centering % Centra la figura orizzontalmente
\hspace*{-0.7cm}\includegraphics[width=0.57\textwidth]{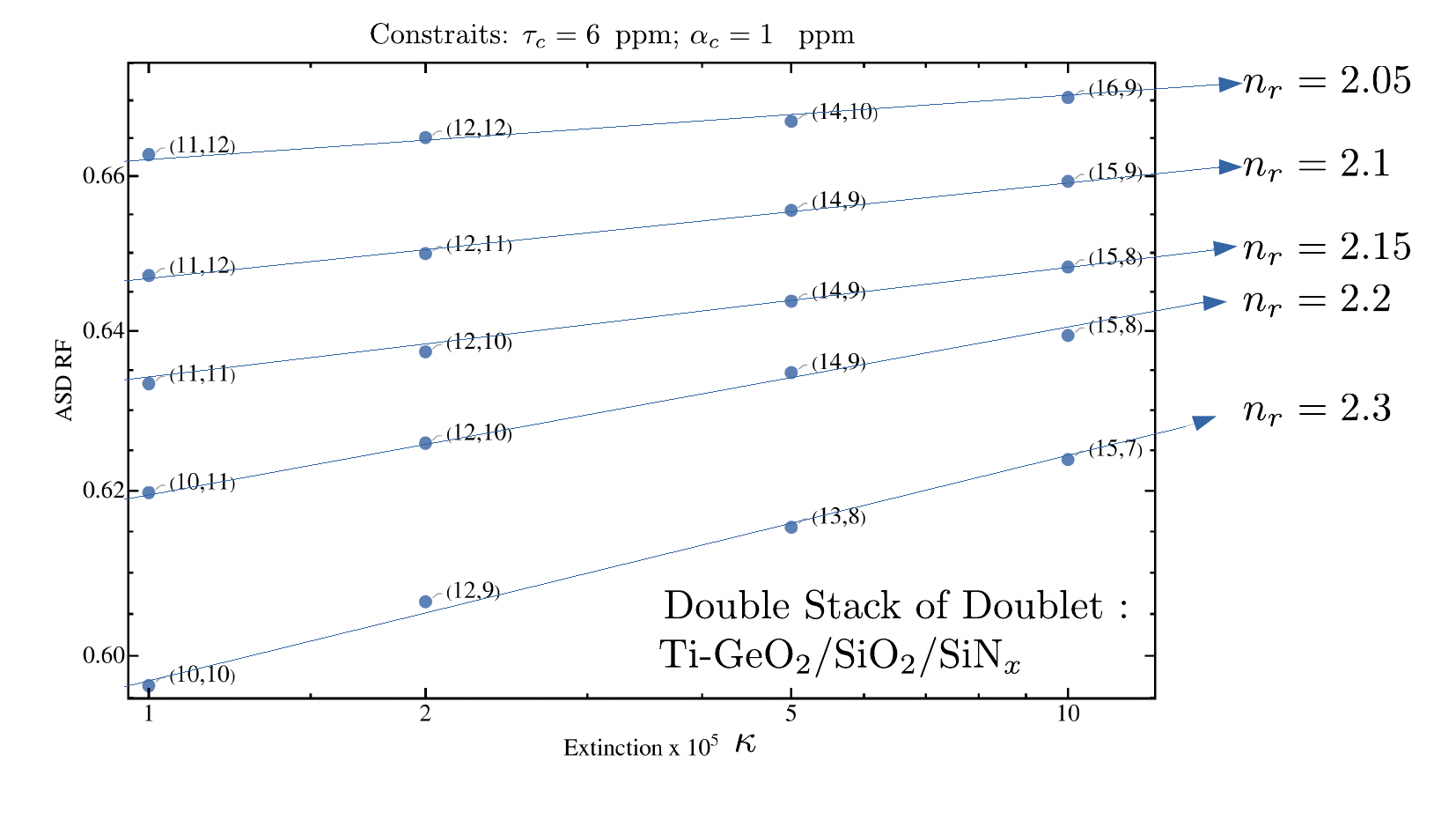} 
\caption{Achievable ASD Reduction Factor (ASD RF) versus the extinction coefficient ($\kappa$) of silicon nitride (SiN$_x$) used as the third material in a ternary DSD structure with Ti::GeO$_2$ and SiO$_2$. 
Each curve corresponds to a different refractive index ($n_r$) for SiN$_x$, ranging from 2.05 to 2.3, in addition to $\eta_{SiN_x}=5.44$. 
Results were generated using a faster optimization algorithm under the $\alpha_c = 1$ ppm, $\tau_c= 6$ ppm constraints. 
The data points are labeled with structure parameters (e.g., number of doublets SiO$_2$/Ti::Ge$_2$ and SiO$_2$/SiN$_x$).}
\label{fig:fig16}
\end{figure}

\begin{figure}[h] % Opzioni di posizionamento: qui, top, bottom, pagina di floats
\centering % Centra la figura orizzontalmente
\includegraphics[width=0.52\textwidth]{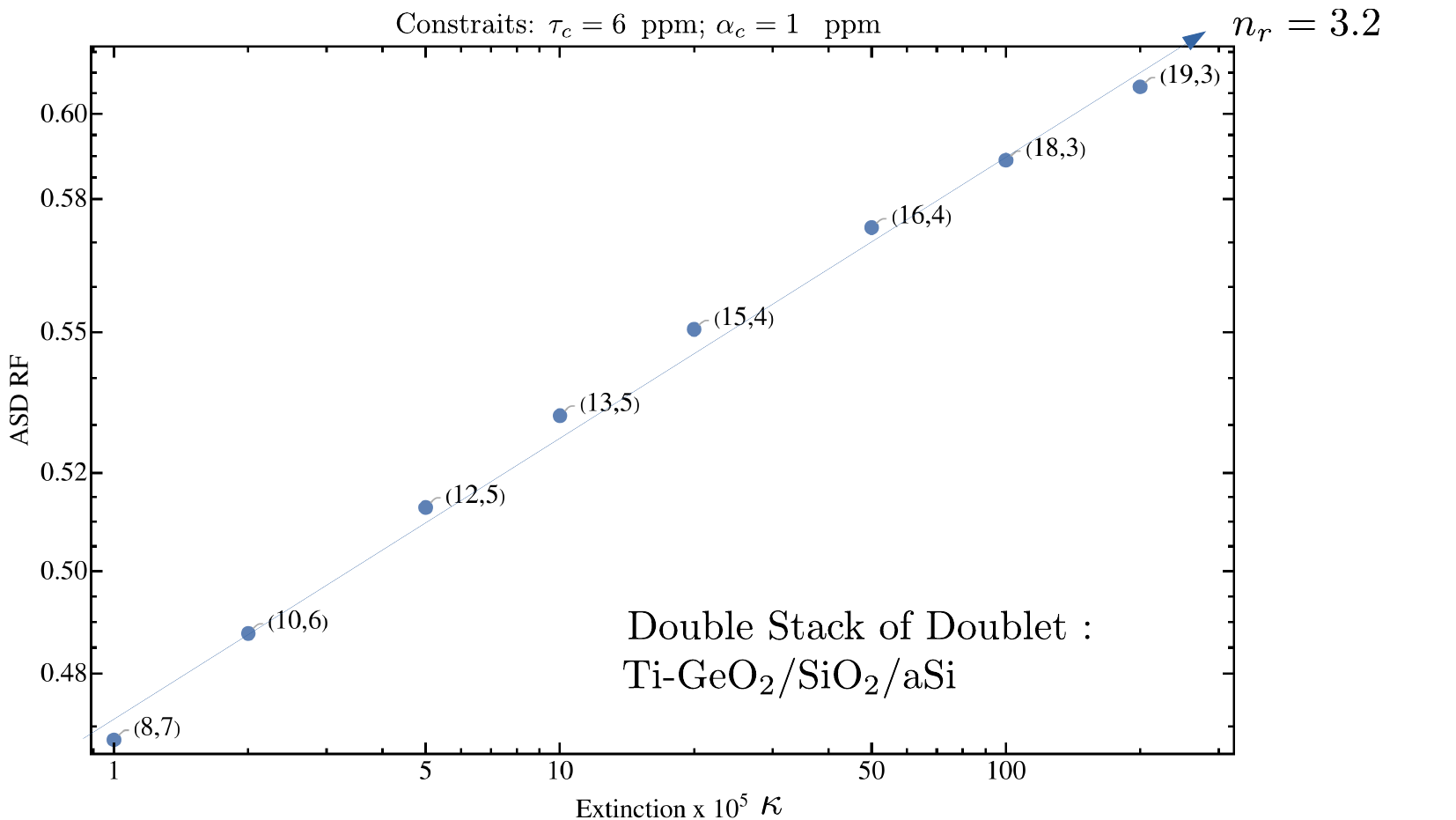} % Includi l'immagine (sostituisci con il tuo file)
\caption{Achievable ASD Reduction Factor (ASD RF) as a function of the extinction coefficient ($\kappa$) of amorphous silicon (aSi, $n_r \sim 3.2$, and $\eta_{aSi}\sim 3.5$) used as the third material in a 
ternary DSD structure with Ti-GeO$_2$ ($n_r=1.89$) and SiO$_2$ ($n_r=1.45$). 
Results obtained using a faster optimization algorithm under the constraints $\alpha_c = 1$ ppm and $\tau_c= 6$ ppm. The data points, labeled with structure parameters
 (e.g., number of doublets SiO$_2$/Ti::Ge$_2$ and SiO$_2$/aSi), 
 indicate that ASD RF values below the 0.5 target are potentially reachable with low-loss aSi.}
\label{fig:fig17}
\end{figure}

Figure \ref{fig:fig16} presents the results for a system using silicon nitride (SiN$_x$) as the third material. 
This figure explores its performance across a range of possible refractive indices ($n_r$ from 2.05 to 2.3) and extinction coefficients ($\kappa$), providing insight into how ASD RF varies with these parameters for SiN$_x$ in this ternary DSD context.
Figure \ref{fig:fig17} subsequently shows the analysis for a ternary system incorporating amorphous silicon (aSi), a material notable for its very high refractive index ($n_r \sim 3.2$). 
Similar to the previous figure, it plots the achievable ASD RF as a function of the extinction coefficient ($\kappa$) of the aSi layer, highlighting the potential impact of this particularly high-index choice on reaching lower ASD RF values.

This analysis demonstrates that by utilizing a high-index material like aSi, it is indeed possible to achieve ASD Reduction Factors below the challenging design target of 0.5, especially when the extinction coefficient 
of the amorphous silicon is kept sufficiently low (e.g., below $\sim 5 \times 10^{-5}$). 
This marks a critical step forward, indicating a viable material pathway towards the desired noise performance.
Furthermore, it is notable that even with relatively high extinction coefficients for aSi, on the order of $\kappa \sim 10^{-3}$ (corresponding to the rightmost points on the plot),
the DSD optimization still yields interesting ASD RF values (around 0.6). 
Significantly, these substantial reductions at higher extinction levels are achieved by using a few pairs of layers or doublets near the bottom of the coating, as indicated by the structural parameters 
labeling the data points (see both Fig.s \ref{fig:fig16} and \ref{fig:fig17} ).
 This suggests that even high-loss, high-index materials could potentially offer competitive performance in specific structures optimized by using a few layers of these materials at the bottom of the coating 
(the result is in line with what was also found in \cite{PierroTernary}). 
 Overall, the incorporation of dense, high-index materials such as aSi or SiN$_x$ into DSD structures represents a promising and effective strategy for approaching and potentially exceeding the ASD RF target of $0.5$.

\section{Conclusions}

This study provided an updated overview of the design challenges and potential pathways for developing low-thermal-noise, high-reflectivity optical coatings, crucial for next-generation gravitational wave detectors and other precision optical systems.
Our focus was on minimizing the ASD RF, with a target goal of 0.5, primarily through the DSD design strategy applied to ternary material systems.
Our analysis began by revisiting the fundamental noise models (Braginsky, Fejer) and highlighting the critical role of material properties, particularly the Braginsky coefficient ($\eta$), extinction coefficient ($\kappa$), and elastic parameters ($Y$, $\sigma$).
We noted significant uncertainty in the determination of $\eta$ for materials like Ti::SiO$_2$, underscoring the need for precise material characterization.
%**
% Indeed, some simulations presented in the Appendix B, examining scenarios with different materials parameters (extreme cases), 
% further underscored the sensitivity of ternary design to the assumed properties, particularly for Ti::SiO$_2$.
%** 

Investigations into binary SiO$_2$/Ti::SiO$_2$ Quarter-Wave Layer (QWL) designs confirmed that while extremely low extinction coefficients $\kappa \sim 10^{-7}$
are necessary to meet stringent absorption requirements (e.g., 0.5 ppm), the achievable ASD RF ($\sim 0.78$) remained far from the target, limited by the mechanical losses described by $\eta$.

We then explored ternary DSD structures based on SiO$_2$ / Ti::SiO$_2$ / Ti::GeO$_2$.
Optimization efforts, constrained by a target absorbance $\alpha_c = 0.5$ ppm, yielded a best ASD RF of approximately $0.698$.
While representing a significant improvement over standard binary designs and establishing a benchmark for this material set, this result fell short of the 0.5 target.
Relaxing the absorbance constraint to $\alpha_c = 1$ ppm led the optimization to favor a binary SiO$_2$/Ti::GeO$_2$ structure, marginally improving the best ASD RF
to $\sim 0.689$, but still not reaching the goal.
Crucially, neither optimization scenario allowed us to reach the target design goal of ASD RF = 0.5. 
Even when permitting absorbance levels as high as 1 ppm, the fundamental properties of this specific material combination,
coupled with the DSD design approach, appear insufficient to simultaneously achieve the desired low coating thermal noise (represented by ASD RF = 0.5) and the imposed absorbance limits.
This strongly suggests that reaching the ambitious 0.5 target will require exploring alternative strategies, such as investigating other materials.

Recognizing these limitations, we shifted our strategy to incorporate denser, higher-refractive-index materials into the ternary DSD design (SiO$_2$ / Ti::GeO$_2$ / H$_2$).
Using a revised, faster optimization algorithm, we evaluated SiN$_x$ and amorphous silicon (aSi) as the H$_2$ component.

Although SiN$_x$ shows notable performance, within the explored parameter range, it doesn't currently seem to emerge as the primary candidate in the strategy aimed at reaching the target.
Instead, the results with amorphous silicon (a-Si, $n_r \sim 3.2$) were highly encouraging.
Our simulations demonstrated that ASD RF values below the $0.5$ target are potentially achievable, provided the extinction coefficient of aSi can be realized at sufficiently low levels ($\kappa \le 5\times 10^{-5}$).
Furthermore, even with higher extinction ($\kappa \sim 10^{-3}$), interesting ASD RF values ($\sim 0.6$) were obtained, suggesting potential utility even for non-ideal material properties.

It is important to acknowledge, however, that the shift towards higher refractive indices and potentially tolerating higher extinction coefficients (even if meeting a relaxed overall absorbance constraint like $1$ ppm)
may necessitate a re-evaluation of auxiliary system requirements.
Specifically, the resulting thermal lensing effects might differ, potentially requiring adjustments to thermal compensation systems, and the optical performance at secondary wavelengths,
 such as the 532 nm band crucial for interferometer alignment, would need careful assessment.

In conclusion, while the SiO$_2$/Ti::SiO$_2$/Ti::GeO$_2$ system offers performance improvements, reaching the ambitious ASD RF target of 0.5 likely requires the integration of advanced,
high-index materials like amorphous silicon. This study highlights a-Si as a promising candidate, contingent on achieving low optical absorption during deposition.
Future work should focus on the experimental realization and characterization of these high-index coatings and potentially explore alternative design paradigms beyond the DSD structure to further
optimize thermal noise performance.

%================================================================================================================================

\appendix
\section*{Appendix: Effective Medium Theory}

The Fejer Effective Medium Theory provides a more general and potentially more accurate model for calculating thermal noise in optical coatings, especially for multilayer structures. It extends beyond the simpler Braginsky model by considering both bulk and shear loss mechanisms within the coating materials.  The general formula for the power spectral density $S_{CB}(f)$ according to the Fejer model is given by:
\begin{widetext}
\begin{equation}
\begin{aligned}
S_{CB}(f) = \frac{2k_B T}{\pi^2 f w_m^2} d \Big\{ \frac{1}{3} \left\langle \frac{1 - 2\sigma}{(1 - \sigma)^2} Y \phi_\perp \right\rangle + \frac{2}{3} \left\langle \frac{1 - \sigma + \sigma^2}{(1 - \sigma) (1 - \sigma^2)} Y \phi_\parallel \right\rangle \Big\} \frac{(1 + \sigma_s)^2 (1 - 2\sigma_s)^2}{Y_s^2} \\
+ \frac{2}{3} \left\langle \frac{(1 - 2\sigma)(1 + \sigma)}{(1 - \sigma)^2} (\phi_\perp - \phi_\parallel) \right\rangle \frac{(1 + \sigma_s)(1 - 2\sigma_s)}{Y_s} \\
+ \frac{1}{3} \left\langle \frac{(1 + \sigma)^2 (1 - 2\sigma) }{(1 - \sigma)^2} \frac{\phi_\perp}{Y} \right\rangle + \frac{2}{3} \left\langle \frac{(1 + \sigma)(1 - 2\sigma)^2 }{(1 - \sigma)^2} \frac{\phi_\parallel}{Y} \right\rangle \Big\}
\end{aligned}
\end{equation}
\end{widetext}
In this formula, the angle brackets $\langle \dots \rangle$ denote a volumetric average over the coating layers. For a two-material coating composed of materials A and B, the volumetric average of a quantity $g$ is calculated as:
\begin{equation}
    \langle g \rangle = \frac{d_A}{d_A + d_B} g_A + \frac{d_B}{d_A + d_B} g_B
\end{equation}
where $d_A$ and $d_B$ are the thicknesses of layers A and B, and $g_A$ and $g_B$ are the values of the quantity $g$ for materials A and B, respectively.

Here, $\phi_\perp$ represents the bulk loss angle and $\phi_\parallel$ represents the shear loss angle.  
The other parameters are as previously defined (temperature $T$, frequency $f$, beam size related to $w_m$, coating thickness $d$, Poisson ratio $\sigma$ and Young's modulus $Y$ 
for the coating, and Poisson ratio $\sigma_s$ and Young's modulus $Y_s$ for the substrate).

A significant challenge in applying the Fejer model lies in the experimental determination of the bulk ($\phi_\perp$) and shear ($\phi_\parallel$) loss angles separately. 
 Traditional measurement techniques often provide a combined or effective loss value.  In practice, and due to the difficulty in independently measuring $\phi_\perp$ and $\phi_\parallel$ \cite{BeS1,BeS2},
  it is frequently assumed that they are approximately equal ($\phi_\perp \approx \phi_\parallel \approx \phi$). 
  This simplification reduces the complexity of the Fejer formula and allows for its practical application using commonly available material loss measurements.  Interestingly, 
  if we assume $\phi_\perp = \phi_\parallel = \phi$, the Fejer formula simplifies, and under certain assumptions regarding substrate contributions, it can be shown to be closely related to the Braginsky model, 
  highlighting the Braginsky model as a specific case of the more general Fejer theory when bulk and shear losses are considered equivalent. However, it is important to acknowledge that the approximation
   $\phi_\perp \approx \phi_\parallel$ might limit the accuracy of the model, especially in materials where bulk and shear losses differ significantly.

There are also more accurate model for coting thermal noise \cite{GV, Hong}. Despite their potential for higher physical accuracy, complex thermal noise coating models often fail 
to significantly outperform simpler ones in fitting data, largely due to experimental uncertainties masking (at present) the finer details.

%================================================================================================================================


\begin{thebibliography}{99}
\bibitem{Nota} the terms "transmittance" and "transmissivity" will be used synonymously, both denoting $1 - |\Gamma|^2$, where $\Gamma$ is the reflection coefficient at the vacuum-coating interface. 
Similarly, "absorbance" and "absorption" will be treated as equivalent terms.
\end{thebibliography}
\end{document}